\documentclass[acmsmall]{acmart}

\usepackage{MnSymbol}
\usepackage[linesnumbered,ruled,vlined,algo2e]{algorithm2e}
\usepackage{algpseudocode}
\usepackage{algorithm}
\usepackage{textgreek}
\usepackage{tablefootnote}
\usepackage{multirow}
\usepackage{overpic}
\usepackage{subfigure}
\usepackage{enumitem}
\usepackage{bm}
\usepackage{xcolor}
\usepackage{tabu}
\usepackage{amsmath}
\usepackage[normalem]{ulem}

\newcommand\graphdb{\textsc{Aster}}
\newcommand\graphkv{\textsc{Poly-LSM}}
\newcommand\graphlsm{\textsc{Poly-LSM}}

\newcommand\graphdbEB{\textsc{Edge-LSM}}
\newcommand\graphdbEBU{\textsc{Delta-Poly}}
\newcommand\graphdbVB{\textsc{Vertex-LSM}}
\newcommand\graphdbVBU{\textsc{Pivot-Poly}}
\newcommand{\etal}{\textit{et al.}}

\definecolor{ao(english)}{rgb}{0.0, 0.5, 0.0}
\newcounter{dingheng}
\numberwithin{dingheng}{section}

\newcounter{fan}
\numberwithin{fan}{section}

\newcounter{junfeng}
\numberwithin{junfeng}{section}

    \newcounter{Siqiang}
\numberwithin{Siqiang}{section}

\setcopyright{rightsretained}
\acmJournal{PACMMOD}
\acmYear{2025} \acmVolume{3} \acmNumber{1 (SIGMOD)} \acmArticle{12} \acmMonth{2} \acmPrice{}\acmDOI{10.1145/3709662}

\ccsdesc[500]{Information systems~Graph-based database models}
\ccsdesc[500]{Information systems~Data management systems}
\ccsdesc[500]{Information systems~Information storage systems}
\ccsdesc[500]{Theory of computation~Data structures design and analysis}
\received{July 2024}
\received[revised]{September 2024}
\received[accepted]{November 2024}

\begin{document}

\title{\graphdb: Enhancing LSM-structures for Scalable Graph Database}

\author{Dingheng Mo}
\affiliation{%
  \institution{Nanyang Technological University}
  \country{Singapore}
  }
\email{dingheng001@e.ntu.edu.sg}

\author{Junfeng Liu}
\affiliation{%
  \institution{Nanyang Technological University}
  \country{Singapore}
  }
\email{junfeng001@e.ntu.edu.sg}

\author{Fan Wang}
\affiliation{%
  \institution{Nanyang Technological University}
  \country{Singapore}
  }
\email{fan008@e.ntu.edu.sg}

\author{Siqiang Luo}
\authornote{Corresponding Author}
\affiliation{%
  \institution{Nanyang Technological University}
  \country{Singapore}
  }
\email{siqiang.luo@ntu.edu.sg}


\begin{abstract}
There is a proliferation of applications requiring the management of large-scale, evolving graphs under workloads with intensive graph updates and lookups.
Driven by this challenge, we introduce {\graphkv}, a high-performance key-value storage engine for graphs with the following novel techniques: 
(1) {\graphkv} is embedded with a new design of graph-oriented LSM-tree structure that features a hybrid storage model for concisely and effectively storing graph data.
(2) {\graphkv} utilizes an adaptive mechanism to handle edge insertions and deletions on graphs with optimized I/O efficiency.
(3) {\graphkv} exploits the skewness of graph data to encode the key-value entries. 
Building upon this foundation, we further implement {\graphdb}, a robust and versatile graph database that supports Gremlin query language facilitating various graph applications.
In our experiments, we compared {\graphdb} against several mainstream real-world graph databases. The results demonstrate that {\graphdb} outperforms all baseline graph databases, especially on large-scale graphs. Notably, on the billion-scale Twitter graph dataset, {\graphdb} achieves up to 17x throughput improvement compared to the best-performing baseline graph system.
\end{abstract}

\maketitle

\section{Introduction}
\label{sec: intro}
Graph database systems provide robust platforms for storing, processing, and analyzing graph data, supporting a variety of applications including social networks, recommendation systems, and biological networks. 
These platforms necessitate rapid responses for online graph navigations where only small fractions of the graph are explored, including intensive data updates (e.g., adding a new edge) and massive data queries (e.g., navigating edges of a vertex)~\cite{fan2022big,davoudian2018survey,microbenchmark}.
For instance, social media users often interact with new connections, while the system must regularly recommend content tailored to their latest activities.  
In such scenarios, both lookup and update efficiencies are critical. Therefore, developing efficient graph database systems for evolving graphs is indispensable for modern applications. 

Existing graph database systems can be {\it in-memory}~\cite{janus,csr,shi2024spruce,fuchs2022sortledton,de2021teseo} or {\it disk-resident}~\cite{Neo4j,orientdb,arangodb,dgraph,sqlg}. In-memory graph database systems consider that the whole graph data resides in main memory, focusing on designing efficient in-memory data structures. Disk-resident graph database systems, in contrast, operate under the assumption that the graph size may exceed the capacity of the main memory, or only a portion of the main memory is available for managing graph data. As a result, these databases primarily store graph data on disk, presenting unique challenges in designing efficient disk-based graph operations. Regarding scalability and data persistence, disk-resident graph database systems such as Neo4j~\cite{Neo4j}, OritentDB~\cite{orientdb}, and DGraph~\cite{dgraph} occupy a large share of the market in the industry. Therefore, this paper aims to develop a comprehensive disk-resident graph database capable of efficiently managing large, evolving graph structures.

\vspace{1mm}
\noindent
{\bf Challenges and limitations of existing graph database systems.} Despite the recent advancements, existing graph database systems still struggle to simultaneously provide strong performance for fundamental graph operators such as querying incident edges of a vertex or updating an edge. These core operations are essential for supporting other important graph queries. 
To facilitate the discussion, let us consider three evaluation criteria: (1) Update, which measures the ability to handle graph evolution, such as edge inserts or deletes; (2) Lookup, which assesses the ability to perform graph traversal along the edges; and 
(3) Scalability, which evaluates the extent of performance degradation as the graph size increases.\footnote{Note that this term does not pertain to the ability of the graph database to scale across multiple machines.}

\begin{figure}[t]
    \centering
    \includegraphics[width=0.80\linewidth]{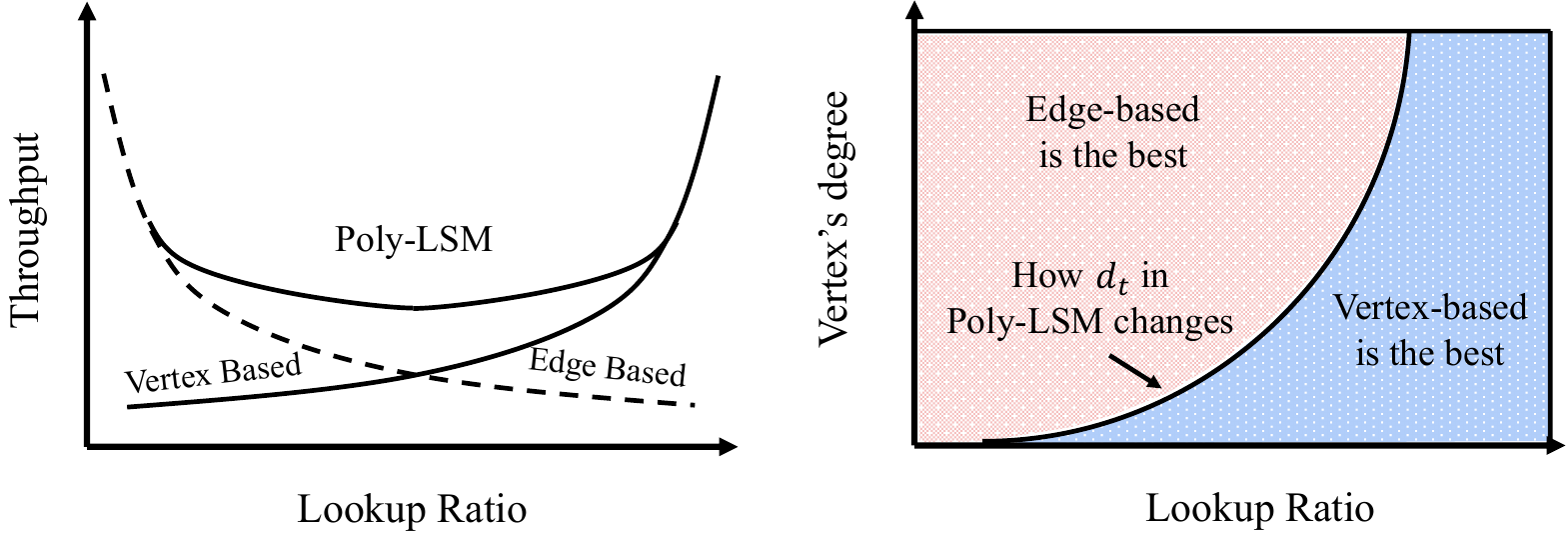}
    \vspace{-2mm}
    \caption{The left figure illustrates the sub-optimal trade-off between lookup and update in existing LSM-based structures. 
    The right figure demonstrates our insight that the superiority of vertex-based updates and edge-based updates are affected by lookup ratio and vertex degree.}
    \vspace{-3mm}
    \label{fig:intro}
\end{figure}

There are three types of mainstream storage engines supporting current (disk-resident) graph database systems, which are {\it linked-list-based}, {\it relational-table-based} and {\it LSM-structure-based}. Neo4j~\cite{Neo4j} and OrientDB~\cite{orientdb} are representatives that employ linked list-based storage, where the adjacency list of a vertex is maintained as a linked list. This approach allows efficient updates, as new edges can be appended directly to the end of the edge file. However, it also leads to inferior lookup performance and unsatisfactory scalability due to incurring substantial random access. In the worst case, a total of $d$ (the degree of a vertex) I/Os is required.
Relational-table-based databases, such as SQLG~\cite{sqlg}, store edges and vertices in separate tables, facilitating efficient updates by appending new entries directly to the tables. However, graph traversal requires costly table joins, limiting neighbor retrieval performance and scalability. 
Log-Structured-Merge (LSM)-tree-based databases such as DGraph~\cite{dgraph} and Sparksee~\cite{Sparksee} can store either an adjacency list (vertex-based) or an edge (edge-based) as a key-value entry (See Section~\ref{sec: basic layouts}). Edge-based LSM-trees offer exceptional update performance and moderate traversal performance by buffering and sequentially flushing new edges that lead to reduced I/Os. Vertex-based LSM-trees, while having strong traversal performance, exhibit lower update performance due to the read-and-modify scheme. 
The analysis of these structures is concluded in Table ~\ref{tab:structure_analysis}.

\begin{table}[t]
\setlength{\abovecaptionskip}{-0.5cm}
\setlength{\belowcaptionskip}{0cm}
\renewcommand{\arraystretch}{1.25}
\setlength{\tabcolsep}{3.5pt}
\scalebox{0.9}{
\begin{tabular}{|c|c|c|c|c|}
\hline
\multicolumn{1}{|l|}{\textbf{Structures}} &
  \begin{tabular}[c]{@{}c@{}}{ Linked} \\ { List}\end{tabular} &
  \begin{tabular}[c]{@{}c@{}}{ Relational} \\ { Table}\end{tabular} &
  \begin{tabular}[c]{@{}c@{}}{ LSM-tree}\\ { (vertex-based)}\end{tabular} &
  \begin{tabular}[c]{@{}c@{}}{ LSM-tree}\\ { (edge-based)}\end{tabular} \\ \hline
 Update &
  \scalebox{0.8}{$\bigstar\bigstar\bigstar$} &
  \scalebox{0.8}{$\bigstar\bigstar\bigstar$} &
  \scalebox{0.8}{$\bigstar\bigstar$} &
  \scalebox{0.8}{$\bigstar\bigstar\bigstar\bigstar\bigstar$} \\ \hline
 Lookup &
  \scalebox{0.8}{$\bigstar\bigstar$} &
  \scalebox{0.8}{$\bigstar$} &
  \scalebox{0.8}{$\bigstar\bigstar\bigstar\bigstar$} &
  \scalebox{0.8}{$\bigstar\bigstar\bigstar$} \\ \hline
Scalability &
  \scalebox{0.8}{$\bigstar\bigstar$} &
  \scalebox{0.8}{$\bigstar\bigstar$} &
  \scalebox{0.8}{$\bigstar\bigstar\bigstar\bigstar$}&
  \scalebox{0.8}{$\bigstar\bigstar\bigstar\bigstar$} \\ \hline
\end{tabular}
}
\vspace{1mm}
\captionof{table}{Summary of different disk-based storage structures.}
\vspace{-1mm}
\label{tab:structure_analysis}
\end{table}

\vspace{1mm}
\noindent
{\bf The Problem: Is there a data structure supporting graph storage with even stronger performance on lookup and update, while achieving high scalability?} From Table~\ref{tab:structure_analysis}, LSM-trees are close to this goal but fall short in either lookup or update performance. A deeper investigation shows that vertex degree plays a crucial factor in determining the benefits of employing vertex-based or edge-based LSM-trees for operations related to a vertex (See Section~\ref{sec: adaptive}).  
Inspired by this observation, we propose a vertex-wise adaptive update mechanism to harness the strength of both vertex-based and edge-based LSM-trees for graph storage, forming a novel graph storage engine named {\graphkv}. {\graphkv} achieves efficient update and lookup operations concurrently for large-scale evolving graphs,  by wisely adapting its LSM-tree update approach to each individual operation.
Poly-LSM supports large-scale evolving graphs with efficient update and lookup performance by wisely adapting its LSM update approach to each individual operation.
Our experiments show that {\graphkv} can achieve up to 17x throughput improvement compared to the state-of-the-art graph database systems. Our novel designs are summarized below.

\vspace{1mm}
\noindent\textbf{Design 1: Multi-layout graph storage to accelerate update and lookup operations concurrently.} 
{\graphkv} employs an innovative LSM-tree-based structure that allows the co-existence of both vertex-based and edge-based graph storage. This design significantly expands the concept of key-value entries, allowing each entry to represent a vertex, an edge, or an adjacency list.
Moreover, {\graphlsm} fully exploits the dynamic nature of the LSM-tree.
It initially represents graph updates in an edge-based form, ensuring decent update efficiency and subsequently utilizes the LSM-tree's dynamic merging process to gradually consolidate entries of the same vertex on disk. 
Therefore, entries at larger LSM-tree levels more closely resemble vertex-based representation, which further enhances the efficiency of lookups.
In this manner, {\graphkv} achieves update efficiency comparable to edge-based LSM-trees and lookup efficiency akin to vertex-based LSM-trees simultaneously.

\vspace{1mm}
\noindent\textbf{Design 2: Adaptive edge update strategy to facilitate evolving graph.}
The hybrid storage layout of {\graphlsm} enables {\graphkv} to deal with edge updates through either an edge-based or vertex-based approach flexibly.
As depicted in Figure ~\ref{fig:intro} (Right), we discover that the overhead of updating edges of the two methods varies from the degree of the updated vertex.
In vertex-based storage approaches, updating an edge for a high-degree vertex necessitates the retrieval and rewriting of a large entry, thereby incurring significant overhead. Hence, the edge-based layout is preferred.
Conversely, for low-degree vertices, this additional cost is relatively low, thus opting for an edge-based storage layout is suboptimal as it incurs higher overhead for later lookups.
Hence, we carefully analyzed the costs associated with two distinct update approaches with different vertex degrees. This analysis yields corresponding selection criteria that can be further bolstered by a degree sketch method~\cite{morris1978counting}. This enables {\graphkv} to adjust the storage layout of the edge update adaptively thus consistently delivering desirable overall performance, as shown in Figure ~\ref{fig:intro} (Left).

\vspace{1mm}
\noindent\textbf{Design 3: Comprehensive graph database to support intricate graph applications.}
We developed {\graphdb}, a comprehensive graph database empowered by {\graphkv} as its storage engine, to effectively manage and navigate graph structures for diverse graph applications ranging from fundamental operations to sophisticated queries. {\graphdb} provides a graph query interface utilizing Gremlin~\cite{rodriguez2015gremlin}, a powerful data-flow query language. Moreover, it is able to efficiently parse graph queries into fundamental operations within {\graphkv} which are subsequently assembled into an execution schedule for supporting complex graph queries. 

\vspace{1mm}
\noindent\textbf{Contributions.} In summary, we make the following contributions.
\begin{itemize}
    \item This paper is at the forefront of efforts to optimize LSM-tree-based graph storage. We introduce {\graphlsm}, a fast storage engine for graphs embedded with a graph-oriented LSM-tree-based structure, with the following techniques: (1) We design an effective and flexible LSM-based storage layout that maps diverse graph structures into polymorphic key-value entries in the LSM-tree and supports both vertex-based and edge-based updates. (2) We scrutinize the impacts of employing various update methods and proposing an adaptive update mechanism that achieves better overall performance. Our analysis indicates that Poly-LSM achieves comparatively optimal processing overhead for a series of graph updates under a uniform workload. (3) We exploit the intrinsic characteristics of graph data within LSM-trees to further improve space efficiency through appropriate encoding of entries.
    \vspace{0.5mm}
    \item We develop and implement a complete graph database, {\graphdb}, on top of {\graphkv} and evaluate its performance against multiple mainstream graph database systems widely used in the industry. The result demonstrates that {\graphdb} outperforms all of them over a wide spectrum of workloads.
\end{itemize}
\vspace{1mm}
\section{Background}
\label{sec: background}
\vspace{1mm}
\noindent\textbf{Key-value Paradigm.} 
Key-value store is a non-relational database paradigm that uses an associative mapping to store data. In this paradigm, data is organized as a collection of key-value entries, where each key serves as a unique identifier and is paired with its corresponding information as the value.
Due to its simplicity, the key-value store has the potential to offer better scalability, usability, and efficiency compared to the traditional relational database paradigm, leading to its increasing popularity in the industry.

\begin{figure}[t]
\vspace{-2mm}
    \centering
    \includegraphics[width=0.45\textwidth]{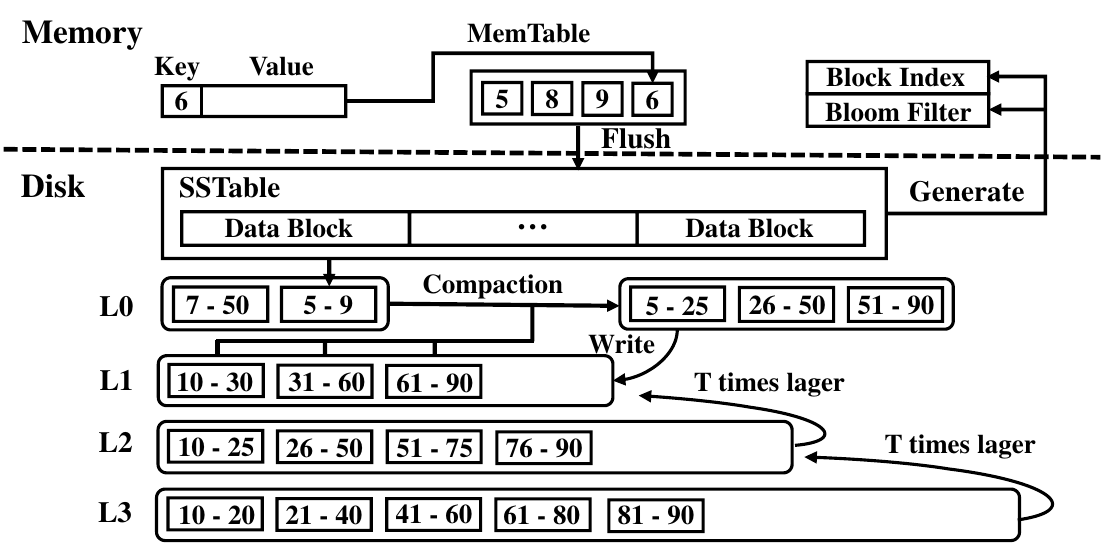}
    \vspace{-3mm}
    \caption{Diagram of the LSM-tree structure.}
    \label{fig:lsmtree}
    \vspace{-2mm}
\end{figure}

\vspace{1mm}
\noindent\textbf{LSM-tree Structure.}  
The LSM-tree is a durable, multi-tiered indexing structure for key-value pairs, which achieves efficient write performance by transforming random writes to the disk into sequential writes.
As illustrated in Figure~\ref{fig:lsmtree}, all updates, insertions, and deletions are initially sorted in a main memory buffer, termed MemTable, as key-value entries. 
When the MemTable reaches its capacity limit, it is converted into a persistent SSTable file stored on the disk.
The SSTables are organized into several levels, with each level having a capacity $T$ times larger than the previous one, and keys are sequential and non-overlapping in SSTables within each level. When the total data size within a level exceeds its capacity, a compaction is triggered. This process involves fetching one or more SSTables from that level and all overlapping SSTables from the next level into memory to perform a merge sort, and then placing the resulting new SSTable into the next level. 

\vspace{1mm}
\noindent\textbf{Lookup in LSM-tree.} The LSM-tree assists queries by indexing the key range and offset of all data blocks in memory and generating Bloom filters for all keys in each SSTable as we depict in Figure~\ref{fig:lsmtree}.
When querying a key, it locates the SSTable containing the key at each level by checking the key range, then uses the Bloom filter to verify the presence of the key. If the Bloom filter returns a positive result, it finds the data block containing the key through the block index and retrieves it from the disk, if it exists. Thus, each lookup incurs at most one block I/O per level, and is very likely to avoid incurring I/O when the key is absent in the level.

\vspace{1mm}
\noindent
{\bf Existing Graph Database Systems.} 
Over the past decades, there are extensive studies on graph database systems that developed many distinctive storage engines. Specifically, Neo4j ~\cite{Neo4j}, one of the most renowned graph database systems, uses linked storage blocks to store vertices and edges respectively. To illustrate, each vertex is formatted as a fixed size entry in the vertex block, where a physical pointer (i.e., position offset in some edge block) is recorded, by which Neo4j can access the edges of a given vertex by doing traversal on the linked list. Similarly, OrientDB ~\cite{orientdb} stores vertices and edges separately and maintains pointers to connect them. However, the pointers in OrientDB are not physical position indicators but rather logical ones, which are mapped to the physical positions in an append-only data structure. Likewise, ArangoDB~\cite{arangodb}, a document-based graph database, serializes each vertex, edge, and the connection of them into a JSON format document. Moreover, some traditional relational databases like PostgreSQL~\cite{momjian2001postgresql} and MySQL~\cite{mysql} also have been adapted, known as SQLG~\cite{sqlg} and MyRocks~\cite{myrocks}, which utilize relational tables to store vertices and edges separately and join them up when doing graph traversal. In addition, JanusGraph ~\cite{janus} stores graph in adjacency list format by using key-value storage backends (i.e., BerkeleyDB by default), where each entry holds the vertex id as the key and the collection of edges of this vertex is the value. Meanwhile, NebulaGraph~\cite{wu2022nebula} serializes each edge or vertex into a key-value entry and stores it in an LSM-tree.

\section{{\graphlsm}: Graph-Oriented LSM-based Storage Engine}
\label{sec:poly_lsm}
This section presents the storage engine {\graphkv}, which is built on an innovative graph-oriented LSM-based design.
\begin{table}[t]
\vspace{0mm}
\centering
\small
\renewcommand\arraystretch{1.02}
  \begin{tabular}{r|l} \toprule
    \textbf{Notation}  & \textbf{Description}  \\
    \midrule
    {$G(V,E)$}              & {The input graph with vertex set $V$ and edge set $E$.}\\
    {$n$}              & {Total number of vertices in the graph.}\\
    {$m$}              & {Total number of edges in the graph.}\\
    {$\overline{d}$}         & {Average degree of the graph.}\\ 
    {$d(u)$}           & {Number of neighbors of vertex $u$.}\\ 
    {$T$}              & {Capacity ratio between adjacent levels in LSM-tree.}\\ 
    {$L$}              & {Number of levels in LSM-tree.}\\ 
    {$B$}              & {Size of a disk block (in bytes).}\\     
    {$I$}              & {Size of a vertex ID (in bytes).}\\        
    \bottomrule
  \end{tabular}
  \vspace{0.5mm}
 \caption{Frequently used notations. }
 \vspace{-6mm}
 \label{table:notation}
 \setlength{\textfloatsep}{0pt}
\end{table}

\vspace{-1mm}
\subsection{Vertex-based and Edge-based Mechanisms}
\label{sec: basic layouts}

We first introduce two typical LSM-tree layouts for storing graphs, which we term as vertex-based LSM-tree and edge-based LSM-tree. 

\vspace{1mm}
\noindent{\textbf{Edge-based LSM-trees.}}
Each graph edge corresponds to a key-value entry in the LSM-tree.
A notable example of graph database systems employing this layout is NebulaGraph, which creates a tuple for each edge and serializes it as a key-value entry for storage.
Hence edge updates simply require inserting a new entry into the LSM-tree. Meanwhile, querying a vertex requires identifying all edges related to that vertex which might access multiple entries at different positions across each LSM level. This is much slower than point lookups even accelerated through range scans.

\vspace{1mm}
\noindent{\textbf{Vertex-based LSM-trees.}}
Each key-value entry represents vertex's adjacency list, where the key represents the vertex and the value contains all its neighbors.
Therefore, querying a vertex \( u \) and its neighbors requires just one fetch of the latest entry keyed by \( u \). 
Nevertheless, inserting new edge $(u,v)$ is more complex, which reads the existing entry containing \( u \)'s adjacency list to update the value, and then reinserted into the LSM-tree. This necessitates an additional lookup for each edge update. In addition, storing all neighbors of a vertex typically leads to large entry size which could introduce significant write cost during the compaction process.

\subsection{{\graphlsm} Layout and Basic Operators}
\label{sec: data model}
We introduce {\graphlsm}, a new LSM-based structure incorporating the merits of both vertex-based and edge-based schemes, to achieve efficient large-scale graph storage on disk. {\graphlsm} supports polymorphic key-value entries and adaptively selects proper update schemes on the fly to enhance system performance.


\vspace{1mm}
\noindent\textbf{Multiple Entry Types.} Different from vertex-based LSM-trees, each vertex in {\graphlsm} has multiple entries to store its neighbor list, including one unique {\it pivot entry} and possibly multiple {\it delta entries}. The pivot entry is usually located at the largest level, with its value containing the majority of the neighbors of that vertex. 
For a vertex, each edge update generates a delta entry containing the updated neighbor list. This delta entry would be merged into the corresponding pivot entry during the subsequent compaction process. In cases where two or multiple delta entries are merged, they would produce another delta entry rather than a pivot entry.

\begin{figure*}[t]
\vspace{0mm}
    \centering
    \includegraphics[width=0.98\textwidth]{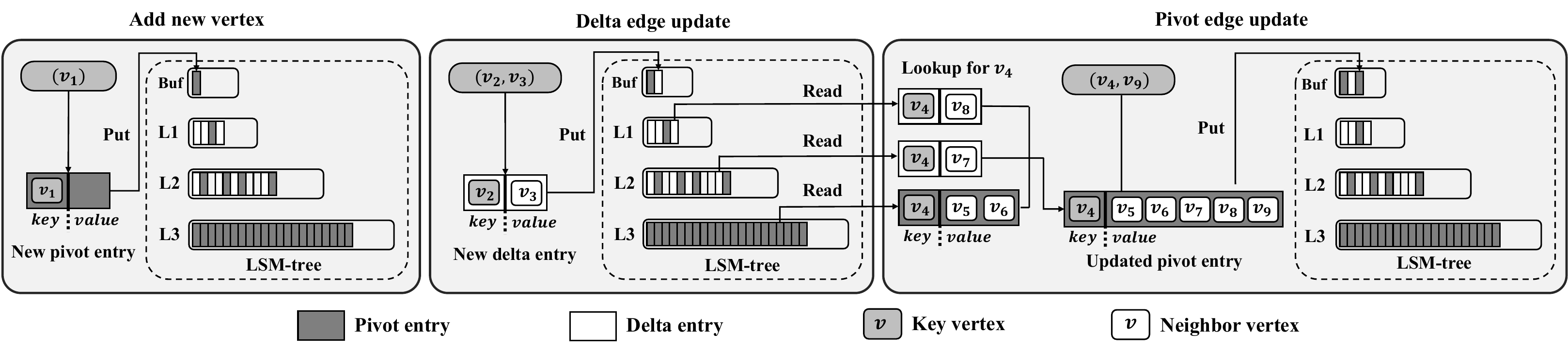}
    \vspace{-1.5mm}
    \caption{An overview of the {\graphkv} demonstrating the workflow of several basic operations including {\it add new vertex}, {\it delta edge update}, and {\it pivot edge update}. In this figure, {\it Buf} represents the memory buffer of the LSM-tree.}
    \label{fig:graphkv}
    \vspace{-3.5mm}
\end{figure*}

\vspace{1mm}
\noindent\textbf{Querying {\graphlsm}.} 
Most complex graph queries and algorithms, such as finding multi-hop contexts and PageRank, can be broken down into a series of adjacency queries of vertices.
These fundamental neighbor retrieval operations in {\graphlsm} are executed by {\it lookup}. Specifically, to get the neighbors of vertex $u$, {\graphlsm} initiates a search from the memory buffer and iteratively searches through subsequent levels for entries with $u$'s ID as the key, and stops until it encounters $u$'s pivot entry.
The values of all relevant entries found are then aggregated to create a comprehensive list of $u$ neighboring edges.
For simpler graph queries, such as inquiries 
a single edge \((u,v)\), {\graphlsm} follows the same processes while returning immediately upon finding the relevant result.


\vspace{1mm}
\noindent\textbf{Updating {\graphlsm}.} 
{\graphlsm} supports both vertex update and edge update. Adding a new vertex can be simply realized by inserting a pivot entry with an empty value, and vertex deletion is achieved with the out-of-place update mechanism in LSM-trees which inserts a special mark of delete.
{\graphlsm} allows two edge updates simultaneously, which benefit different scenarios.

\vspace{1mm}
\noindent\underline{\textit{Delta Edge Update.}} For adding an edge $(u, v)$ in the graph, {\graphlsm} simply needs to insert a delta entry into the LSM-tree with the key as $ u $'s ID and the value as $ v $'s ID. \footnote{Note that in undirected graphs, an edge update $ (u, v) $ is always accompanied by another edge update $ (v, u) $, and the processing of the latter is identical to the former. Therefore, in our subsequent discussion, we will only mention the edge update $ (u, v) $.} In addition, an edge deletion is also performed as the insertion of an entry, where a specific value is designated to signify the deletion of that edge.

\vspace{1mm} 
\noindent\underline{\textit{Pivot Edge Update.}} 
To add an edge \((u, v)\) with pivot update, {\graphlsm} first conducts a lookup for vertex \( u \) to retrieve its current neighboring edges. It then forms a new pivot entry for \( u \), appending the ID of vertex \( v \) to the fetched list of neighboring edges. This updated pivot entry is subsequently inserted into the LSM-tree.  For instance, as shown in Figure~\ref{fig:graphkv}, to add edge \((v_4, v_9)\), {\graphlsm} firstly lookup the two delta entries in Level 1 and Level 2, and the pivot entry in Level 3, with values of \(\{v_8\}\), \(\{v_7\}\), and \(\{v_5, v_6\}\) respectively. These, along with the new neighbor \(v_9\), are merged to form the updated pivot entry value \(\{v_5, v_6, v_7, v_8, v_9\}\), which is then appended to the memory buffer. 
Edge deletions follow a similar process, with the distinction that the edge is to be removed from the lookup results before forming the updated pivot entry.

The delta update increases the number of valid entries corresponding to a vertex, resulting in higher read amplification when querying that vertex. Therefore, it is more suitable for scenarios with a lower lookup ratio in the workload. On the other hand, the pivot update requires reading all the neighbors of the vertex being updated first, making it more suitable for vertices with a low degree, as the total data size that needs to be read is smaller.

\vspace{1mm} 
\noindent\textbf{Hybrid Layout in {\graphlsm}.} In practice, for each edge update, {\graphlsm} would adaptively select the most appropriate approach from these two update methods. This approach is more flexible than the default merge behavior in RocksDB, which relies solely on a fixed delta update method. As a result, it offers greater potential for enhancing system performance. Typically, an edge update is performed as delta updates, while pivot updates are used for vertexes with low degree or under workloads with fewer lookups. We will present detailed reasoning by analyzing the distinct trade-off in these two schemes in Section~\ref{sec: adaptive}.
In this manner, a newly updated edge $(u, v)$ in {\graphlsm} likely first materializes in the edge-based layout and subsequently moves to deeper levels through compaction, eventually merging into the vertex-based pivot entry corresponding to vertex $u$. This hybrid storage layout enables {\graphkv} to be more versatile in managing large, evolving graphs.

\vspace{1mm}
\noindent\textbf{Practical Implementation in RocksDB.}
After discussing the high-level layouts of {\graphlsm}, we will proceed to describe how we extended RocksDB to implement it in practice.

\vspace{0.5mm}
\noindent\textit{\underline{Default RocksDB. }}
By default, RocksDB uses \textit{Put} interface to insert data. However, the inserted entries will overwrite previous entries with the same key during compaction. Algorithmically, this means that each inserted entry acts as a pivot entry. 

\vspace{0.5mm}
\noindent\textit{\underline{Merge Operator. }}
To support delta entries, we leverage the \textit{Merge Operator}~\cite{mergeoperator,cao2020characterizing,xu2024ionia} API provided by RocksDB. This is a user-defined interface that allows specifying custom behaviors for handling multiple values associated with the same key. When entries are inserted using the \textit{Merge} operation, they are combined with other entries of the same key according to the behavior defined by the merge operator, rather than discarding obsolete values like \textit{Put}.

\vspace{0.5mm}
\noindent\textit{\underline{Our Implementation.}} We override the {\it Merge Operator} in RocksDB to realize delta entries in {\graphlsm}'s hybrid layout. The basic \textit{Merge Operator}~\cite{mergeoperator} in RocksDB appends the value of a new entry to those with the same key. Whereas, we significantly enrich the semantics of this interface to support graph storage with following key features. 
(1). Our implementation ensures no duplicate edges within an adjacent list. For example, suppose we have a pivot entry with the key as node $u$ and the value ${v, w}$, and a delta entry with the value ${v}$. The value produced by RocksDB’s default merge operator is ${v, w, v}$, while our implementation gives ${v, w}$ with duplicate edges filtered out. 
(2). Our custom \textit{Merge Operator} ensures edges in the value being ascending sorted by node ID which is not guaranteed by RocksDB's basic appending method. This facilitates subsequent merging, querying, and value encoding.
(3). The \textit{Merge Operator} interface only supports value accumulation, making it unable to handle edge deletions independently. We address this limitation by adding a label to the delta entry's value which triggers the edge removal during the merge process.

With this implementation, hybrid updates can be supported by invoking different RocksDB APIs.
Specifically, pivot updates are handled by \textit{Get} and \textit{Put} interfaces to read existing entries and insert updated version.
While delta updates solely rely on \textit{Merge} interface to append the edge entry. {\graphlsm} employs a carefully designed adaptive algorithm to determine the optimal timing for applying pivot updates versus delta updates, as we will detail in next section.

\subsection{Adaptive Updates in {\graphlsm}: Toward Optimal I/O Efficiency}\label{sec:adaptive}
\label{sec: adaptive}
{\graphkv} would adaptively employ the most appropriate method for each incoming edge update $(u,v)$ based on the current workload and the degree of $u$, which effectively decreases the expected I/O cost. The rationale is that the pivot update incurs additional costs at insertion, as it requires reading and rewriting the existing neighbors of the edge's vertices. Moreover, the updated pivot entry is usually larger than a delta entry, which leads to more I/Os in subsequent compactions. On the other hand, the delta update leads to additional costs during subsequent related queries. This additional cost arises because, until the new edge is merged with the pivot entry of the corresponding vertex during compaction, any query targeting the vertex requires an independent I/O operation to access this newly updated edge. Next, we will analyze the I/O costs associated with the delta and pivot update methods in detail. 
Specifically, in line with prior works~\cite{raju2017pebblesdb,huynh2021endure,dostoevsky2018}, we assume a static workload with fixed proportions of different operations and a uniform key-value distribution. The LSM-tree is assumed to have a balanced structure, where the size ratio between consecutive levels remains \(T\), including the largest level.
The frequently used notations are listed in Table ~\ref{table:notation}.

\vspace{1mm}
\noindent\textbf{Cost of Delta Update.} Equation~\ref{eq: edge-based cost} represents the expected cost of delta update $C_D$ of the edge $(u, v)$. The first term accounts for the cost of writing new delta entry into the LSM-tree. The second estimates the additional prospective read cost incurred for lookups of vertex \(u\) before the delta entry merges into \(u\)'s pivot entry.

\begin{equation}
\label{eq: edge-based cost}
C_{D} = \underbrace{\frac{2I\cdot T \cdot L}{B}}_{\text{Write I/O cost}} + \underbrace{N_L\cdot P_u}_{\substack{\text{Prospective} \\ \text{read I/O cost}}}
\end{equation}

\vspace{-1mm}
\noindent\underline{\textit{Write I/O Cost.}}
In the first term, \(\frac{2I}{B}\) represents the number of block I/Os generated each time the new delta entry is written to disk, where \(I\) is the size of the vertex IDs in entries in bytes, and \(B\) is the size of a disk block in bytes. To determine the actual write I/O cost, this value must be multiplied by the LSM-tree's inherent write amplification, which is the product of the adjacency ratio \(T\) and the total number of levels \(L\).

\vspace{0.5mm}
\noindent\underline{\textit{Prospective Read I/O Cost.}} In the second term,  $ N_L $ indicates the expected number of lookups in the workload that occur from the insertion of the new delta entry until it merges with $u$'s pivot entry and $ P_u$ is the probability that a lookup targets vertex $u$. 
The expression \( N_L \cdot P_u \) represents the number of lookups for vertex \( u \) during the existence of the delta entry corresponding to the new edge \((u, v)\). Each of these lookups incurs an additional block I/O due to the need to fetch this delta entry.
We model the expected number of operations that occurred during the existence of edge $(u, v)$'s delta entry in Equation~\ref{eq: W}, where $m$ is the total number of edges in the graph, $\theta_L $ and $ \theta_U $ denotes the ratio of lookups and updates in the workload, respectively. 
\begin{equation}
\label{eq: W}
    W=\frac{m \cdot \theta_L}{(T-1)\cdot \theta_U}
\end{equation}
The rationale is that from the moment an entry is newly written to the LSM-tree until it reaches the final level, on average, a number of edges equivalent to the sum of the capacities from the first to the penultimate level are written into the LSM-tree. Since each level is $T$ times larger than its precedent level, this amount roughly constitutes $\frac{1}{T-1}$ of the total size of the LSM-tree, which contains $m$ edges in total. As for $P_u$, because there are $n$ vertices on the graph, and the lookup might target at any vertex, we have $P_u={1}/{n}$.
Finally, 
we can derive the following equation, where $\overline{d} = \frac{m}{n}$ denotes the average degree of the graph.
\begin{equation}
C_D=\frac{2I\cdot T \cdot L}{B} + \frac{\theta_L \cdot \overline{d}}{ \theta_U \cdot (T-1)}
\end{equation}


\vspace{1mm}
\noindent\textbf{Cost of Pivot Update.} The expected cost of adding an edge $(u, v)$ with the pivot update can be written as Equation~\ref{eq: C_P}, which also consists of two terms: (1). The read I/O cost of conducting a query to fetch neighboring edges of vertex $u$ from disk. (2). The cost of rewriting the updated pivot entry into the LSM-tree.
\begin{equation}
\label{eq: C_P}
    C_P = \underbrace{2 + \frac{(d(u)+1)\cdot I}{B}}_{\text{Lookup cost for $u$}} + \underbrace{\frac{(d(u) + 2)\cdot I \cdot T \cdot L}{B}}_{\text{Rewrite I/O cost}}
\end{equation}

\vspace{0.5mm}
\noindent\underline{\textit{Rewrite I/O Cost.}} In the second part of Equation~\ref{eq: C_P}, $(d(u)+2)\cdot I$ represents the size of the updated pivot entry, which includes one vertex ID in the key and $(d(u)+1)$ IDs in the value. We multiply this by the LSM-tree's write amplification \( TL \) then divide by disk block size \( B \) to derive block I/O incurred by writing this entry.

\vspace{0.5mm}
\noindent\underline{\textit{Lookup Cost for $u$.}}  The cost of performing a lookup for vertex $u$ can be further divided into the I/Os that may occur when finding delta entries of $u$ in a level, and the necessary I/Os when reading out $u$'s pivot entry upon its encounter. ~\footnote{During the lookup for \( u \), false positives from Bloom filters might occasionally result in disk I/Os, but this occurrence is negligible. In {\graphkv}, the Bloom filter is configured with 10 bits per key, a standard setting in LSM-based key-value databases. This configuration keeps the false positive rate below 1\%, making it highly unlikely.}

We estimate the expected overhead for retrieving delta entries of vertex $u$ as approximately one I/O considering the significantly reduced size of smaller levels. From previous analysis, the last level holds roughly $\frac{(T-1) \cdot m}{T}$ edges, leading to the $(L-i)$-th level containing $\frac{(T-1) \cdot m}{T^{1+i}}$ edges.
For uniform key distribution, the probability that the target delta entry is found at the $(L-i)$-th level is $P_L^i = 1-(1-\frac{1}{n})^{\frac{(T-1) \cdot m}{T^{1+i}}}$, where $\frac{1}{n}$ denotes the probability of an update operation involving the target vertex. 
Since $m = n \cdot \overline{d}$, we can reformulate $P_L^i$ as $(1-(\frac{n-1}{n})^{n \cdot \frac{(T-1) \cdot \overline{d}}{T^{1+i}}})$. In our context, the total vertex number $n$ is typically large, indicating $(\frac{n-1}{n})^{n}$ approaches $\frac{1}{e}$, which yields the following equation.
\begin{equation}
P_L^i \approx 1 - e^{-\frac{(T-1) \cdot \overline{d}}{T^{1+i}}}
\label{equation: P_L}
\end{equation}
It is clear that as $i$ increases, $P_L^i$ declines sharply, with its value affected by $\overline{d}$ and $T$ as well. For real-world graphs, whose average degrees typically range from 10 to 100, Equation~\ref{equation: P_L} indicates that the probability \( P_L^1 \) of accessing delta entries at the $(L-1)$-th level is close to 1 when RocksDB's default size ratio 10 is employed. 
In contrast, for smaller levels, \( P_L^2 \) and \( P_L^3 \) decrease drastically even approaching zero. 
This allows us to approximate the expected overhead for retrieving delta entries for vertex \( u \) as roughly 1 I/O, due to the minimal contributions from levels smaller than the $(L-1)$-th level. Besides, the precise expected overhead can also be calculated by collecting the expected retrieval count across all levels:
\begin{equation}
C_R = \sum_{i=1}^{L-1} P_L^i.
\end{equation}
More specifically, consider the Wikipedia dataset~\cite{yang2012defining}, whose average degree is 37.11, the probability at the (L-1)-th level is \( P_L^1 = 0.964 \). While at the (L-2)-th level, this probability reduces to \( P_L^2 = 0.284 \), and at the (L-3) level, it diminishes further to \( P_L^3 = 0.033 \).

The expected cost of retrieving the pivot entry equals $1+\frac{(d(u)+1)\cdot I}{B}$. In {\graphkv}, entry sizes are irregular, which can lead entries to span multiple disk pages and consequently induce multiple block I/Os during retrieval. Assuming the starting positions of entries on the disk relative to the start of a block are uniformly random, the expected number of blocks an entry spans is its length \((d(u)+1) \cdot I\) divided by the disk block size \(B\), plus one.

Combining these factors, the total I/O cost of the lookup for $u$ is thus calculated to be $2 + \frac{(d(u)+1)\cdot I}{B}$.

\vspace{1mm}
\noindent\textbf{Adaptive Selection.}
The delta update incurs lower costs compared with the pivot update when \(C_P > C_D\), which can be written as
\begin{equation}
\label{eq: adaptive threshold}
2 + \frac{(d(u)+1)\cdot I}{B} + \frac{d(u)\cdot I\cdot T \cdot L}{B}> \frac{\theta_L \cdot \overline{d}}{\theta_U \cdot (T-1)}
\end{equation}
Equation~\ref{eq: adaptive threshold} gives a threshold \(d_t\) expressed in Equation~\ref{eq: d_t}. When \(d(u)\geq d_t\), {\graphkv} always adopts the delta update for edge $(u,v)$. Otherwise, {\graphkv} employs the pivot update.  

\begin{equation}
\label{eq: d_t}
d_t =\max\{ \lceil \frac{\theta_L \cdot \overline{d} \cdot B}{\theta_U \cdot  I \cdot (T-1) \cdot (T \cdot L+1)} - \frac{2B}{I\cdot(T \cdot L+1)} - \frac{1}{T \cdot L+1} \rceil , 0\}
\end{equation}

Here we give a practical running example. 
We set the vertices and edges as 64-bit integers, each occupying 8 bytes, which gives $I=\text{8 Bytes}$. We let the disk block size $B$ be 4KB, the LSM-tree's adjacency ratio $T$ be 10, the number of levels in LSM-tree $L$ be 4, and the average degree of input graph $\overline{d}$ be 32, which all reflect typical real-world configurations.
We assume the workload comprises equal proportions of lookups and updates, resulting in  $\theta_L=0.5$, and $\theta_U=0.5$. 
Then, as specified by Equation~\ref{eq: d_t}, we have $d_t = 21$. This means that in this scenario {\graphkv} opts for the delta update when the degree of the source vertex is 21 or higher. For lower degrees, it employs the pivot update.

According to Equation~\ref{eq: d_t}, when the workload is update-intensive, the parameter \(d_t\) becomes very small. This causes the majority of vertex updates to be handled as delta updates, which enhances write performance. Conversely, when the workload is lookup-intensive, \(d_t\) becomes very large, leading most updates to be processed as pivot updates, thereby improving read performance. 
This adaptability enables {\graphlsm} to provide solid throughput in any scenario, offering better robustness compared to other graph database systems. 

\noindent\textbf{Is the Adaptive Update Scheme Optimal?} 
For an evolving graph \(G(V, E)\), consider a sequence of $N$ edge updates, where each update can be chosen between a delta update or a pivot update. Essentially, there are \(2^N\) possible update sequences, and we denote \(S^*=\{\nu^*_1, \nu^*_2 ... \nu^*_N\}\) as the optimal sequence which incurs the smallest overhead among all possible combinations. Further let \(S=\{\nu_1, \nu_2 ... \nu_N\}\) represent the sequence chosen by {\graphlsm}. {\it We can show that the cost of $S$ is very close to that of the optimum $S^*$.} In particular, when the workload is uniform across graph vertices, \(S\) is exactly the optimal sequence \(S^*\). In more complicated situations, when the workload is skewed over graph vertices, the cost of \(S\) is at most a $\log m$ factor of \(S^*\), as Lemma~\ref{lemma: optimality} indicates.\footnote{The full proof of Lemma~\ref{lemma: optimality} is included in our technical report~\cite{technicalreport}.}

\begin{lemma}
When the workload is uniformly distributed, the total cost of {\graphlsm}  is $O(1)$-competitive. When the workload is skewed, {\graphlsm} is $O(\log m)$-competitive.
\label{lemma: optimality}
\end{lemma}

\noindent\textbf{Extending to 1-Leveling LSM-tree Designs.}
In our analysis, we adopt the commonly used leveling structure~\cite{luo2020breaking, dayanspooky, huynh2021endure, dayan2017monkey} by default, which maintains a single sorted run at each level. However, our approach can be easily extended to other structures through adjustments to the cost model. For example, in practice, RocksDB typically employs a variation known as 1-leveling, where each buffer flush is stored as a separate run in the first level and does not merge with existing data. In this case, the cost of a delta update is given by $\frac{2I\cdot (T(L - 1) + 1)}{B} + \frac{\theta_L \cdot \overline{d}}{ \theta_U \cdot (T-1)}$, where the first term represents the modified write I/O cost for a delta entry. The rationale is that each entry incurs only one time write amplification at the first level and $T$ times at the remaining levels under the 1-leveling design. Correspondingly, the cost of a pivot update is $2 + \frac{(d(u)+1)\cdot I}{B} + \frac{(d(u) + 2)\cdot I \cdot (TL - T + 1)}{B}$.
According to Equation ~\ref{eq: adaptive threshold}, a delta update outperforms a pivot update when 
\begin{equation}
2 + \frac{(d(u)+1)\cdot I}{B} + \frac{d(u)\cdot I\cdot (T \cdot L - T + 1)}{B}> \frac{\theta_L \cdot \overline{d}}{\theta_U \cdot (T-1)}
\end{equation}
This leads to the corresponding threshold \(d_t'\) as presented
\begin{equation}
\label{eq: d_t 1-leveling}
d_t' =\max\{ \lceil \frac{B}{I\cdot(T \cdot L-T+2)} \cdot (\frac{\theta_L \cdot \overline{d}}{\theta_U \cdot (T-1)} - 2) - \frac{1}{T \cdot L-T+2} \rceil , 0\}
\end{equation}
Notably, the threshold with 1-leveling is slightly higher than that of the leveling structure, which suggests an increased likelihood of using pivot updates.


\vspace{1mm}
\noindent\textbf{The Degree Sketch.}
When adding an edge \((u, v)\), {\graphkv}'s adaptive mechanism uses \(u\)'s degree to determine the appropriate edge update method, based on the threshold outlined in Equation 6. Therefore, it is necessary to index the degree information of vertices in memory to avoid I/O overhead during this decision-making process.
However, directly storing degree information in memory for very large graphs consumes significant space. On the other hand, compressing degree information is not viable, as it cannot accommodate changes in degrees caused by intensive edge updates in evolving graphs.
{\graphkv} employs an approximate sketch based on Morris Counter~\cite{morris1978counting} to index the degree of each vertex in memory with minimal space overhead, which also supports dynamic updates to the degree information. 

\begin{algorithm}[b]
  \caption{Update Degree Sketch}
  \label{alg: degree sketch}
  \hspace{-44mm}\KwIn{a new edge $(u, v)$}
  \begin{algorithmic}[1]
    \State {$E_u \gets sketch[u] >> 4$;}
    \State {$r\gets rand(0,1)$;}
    \State{\textbf{if} $r < 2^{-E_u}$}
    \State {\ \ \ \ $sketch[u]\gets sketch[u]+1$;}
    \end{algorithmic}
\end{algorithm}

\begin{figure}[t]
    \centering
    \includegraphics[width=0.5\textwidth]{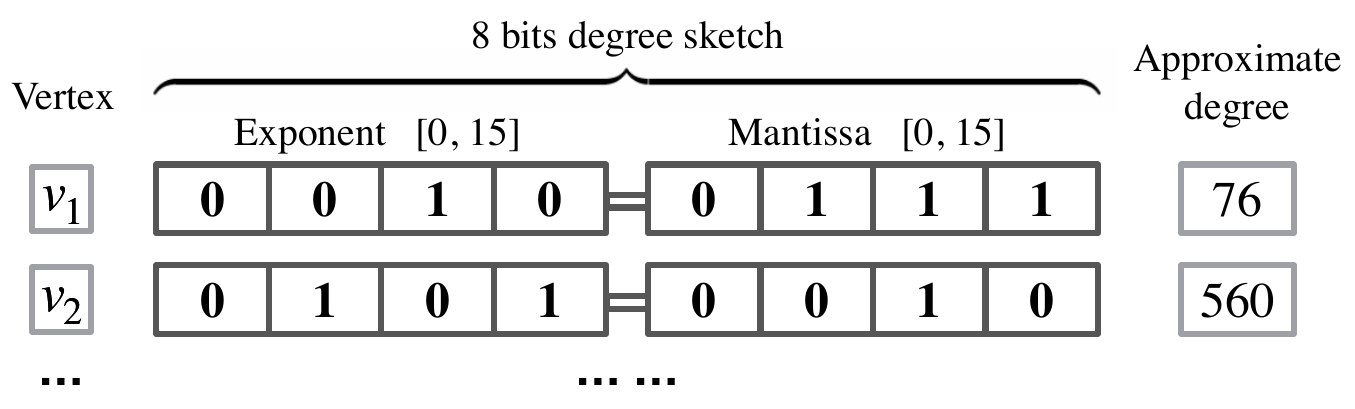}
    \vspace{0mm}
    \caption{The degree sketch allows {\graphkv} to index the degree of each vertex in memory with very few bits.}
    \vspace{-1mm}
    \label{fig:morris}
\end{figure}

The key idea of the degree sketch involves accounting for degree increments based on a probability that correlates with the current degree size. As  Figure~\ref{fig:morris} illustrates, for each vertex, {\graphkv} allocates 8 bits to index its degree in memory, divided into two parts: the first 4 bits termed as the exponent, and the last 4 bits termed as the mantissa, each corresponding to an integer between 0 and 15. The update process of the degree sketch is shown in Algorithm~\ref{alg: degree sketch}. Let ${E}_u$  denote the exponent of vertex $u$ in the degree sketch, and ${M}_u$ denote its mantissa. Every time the degree of vertex $ u $ increments, ${M}_u$ has a probability of $\frac{1}{2^{{E}_u}}$ to increment. Otherwise, ${M}_u$ remains unchanged. If $ M_u $ is incremented upon reaching its maximum value of 15, it will be reset to 0, and then $ E_u $ is increased by 1.
We use the equation below to calculate the approximate value of $u$'s degree.
\begin{equation}
    \hat{d}(u) = (2^{E_u} - 1) \cdot 2^4 + 2^{E_u}\cdot M_u
\end{equation}
For example, in Figure~\ref{fig:morris}, for vertex $v_1$ we have $E_u = 2$ and $M_u = 7$, thus the approximate degree that it sketches is $\hat{d}(v_1) = (2^2-1)\times 2^4 + 2^2\times 7 = 76 $. Similarly, the approximate degree for $v_2$ is $\hat{d}(v_2) = (2^5-1)\times 2^4 + 2^5\times 2 = 560 $.
It is obvious that for any vertex $u$, $\hat{d}(u)$ is an unbiased approximation of $u$'s actual degree, which satisfies $\mathbb{E}[\hat{d}(u)] = d(u)$.
Furthermore, our degree sketch provides a steady approximation for any degree scale, as the following lemma summarizes.
\begin{lemma} 
When a graph $G(V,E)$ is loaded into {\graphkv}, for any $u\in V$, the approximation $\hat{d}(u)$ returned by the degree sketch of {\graphkv} satisfies
\begin{equation}
\label{eq: sketch lemma}
{\rm{Pr}}[|\hat{d}(u)-d(u)|\geq\epsilon d(u)]\leq \frac{1}{6\epsilon^2 }
\end{equation}
\end{lemma}
\renewcommand*{\proofname}{Proof Sketch}
\begin{proof}
According to Corollary 3 in~\cite{csHuros2010approximate}, we have 

\begin{equation}
\label{eq: variance bound of degree sketch}
\frac{{\rm{Var}}[\hat{d}(u)]}{\mathbb{E}[\hat{d}(u)]} \leq \sqrt{\frac{3}{8\cdot 2^4 - 3}} \leq \frac{1}{6}
\end{equation}
Applying Chebyshev's inequality to Eqaution~\ref{eq: variance bound of degree sketch}, we can derive the conclusion in Equation~\ref{eq: sketch lemma}.
\end{proof}

In practice, the relative error of our degree sketch's approximation is around 10\% for any degree. 
This level of precision is adequate for applying adaptive edge updates, as {\graphkv} selects the update method based on an adaptive degree threshold. For vertices with degrees that vary by no more than 10\% from this threshold, the cost of applying either update method is comparable.
Furthermore, the memory consumption of the degree sketch is generally notably lower than that of the Bloom filters. This is because it allocates only 8 bits per vertex, whereas the Bloom filters allocates 10 bits for each key in LSM-tree, which can include duplicate vertices.
Since both the exponent and mantissa are between 0 and 15, the maximum value that the degree sketch can represent is $\hat{d}_{max}=(2^{15}-1)\cdot 2^4 + 2^{15}\cdot 15 = 1015792$, which is sufficiently large.
In instances where a vertex's degree exceeds the maximum limit, {\graphkv} would consistently select the edge-based update for this vertex to avoid the extreme cost of retrieving all its neighbors.




\subsection{Improve Space Efficiency with Partitioned Elias-Fano Encoding}
\label{sec: encoding}

The space efficiency of {\graphkv} could be further enhanced by customizing the encoding method to the specific characteristics of graph storage.
As outlined in Section ~\ref{sec: data model}, each entry in {\graphlsm} consists of a key, representing the ID of a vertex \( u \), and a value, which is a sorted list of vertex IDs detailing \( u \)'s neighboring edge. The vertex IDs are integers constrained by $n$, the total number of vertices in the graph, making the inverted index compression techniques~\cite{yan2009inverted} particularly suitable. Additionally, edge distribution for a vertex is often skewed in real-world applications. Therefore, we propose encoding the entry value using the partitioned Elias-Fano method~\cite{ottaviano2014partitioned}, which provides desirable compression rates and encoding/decoding efficiency by considering edge locality. 

In our implementation, the partitioned Elias-Fano algorithm divides the vertex ID list $\{k_i\}$ in an entry into $t$ segments and encodes the local information using a two-level structure.
Assume each segment contains $s$ consecutive vertex IDs within a sub-universe $N_j$, where $j$ denotes the segment ID. The first level comprises a list of the starting ID of each segment and the terminating ID of the last segment to identify them.
In the second level, each segment is encoded using the Elias-Fano algorithm. Specifically, the sub-universe $N_j$ is further divided into equal-length sub-segments represented by different prefixes. Within each sub-segment, each ID is divided into a prefix and remaining bits. The Elias-Fano algorithm then unary encodes the ID counts concatenated with the remaining bits of each ID, requiring $2 + \log_2\frac{N_i}{t}$ bits per element. Additionally, the ID list in the first level can be encoded similarly, consuming approximately $2 + \log_2 t$ bits per segment.
It is worth mentioning that the compression rate can be adjusted by tuning the number of segments or the prefix length in the Elias-Fano encoding, depending on the distribution of edges.

\subsection{Adapt for Various Graph Format}\label{sec:graphformat}
We will now describe how these previously discussed techniques are applied to directed graphs in our practical implementation. 

Firstly, at the data model level, it is necessary to distinguish between out-edges and in-edges.
Each entry's value is separated into two sorted lists, representing the ID of out-neighbors and in-neighbors respectively. 
For instance, processing with the edge-based update entails inserting two entries: one with a key of $ u $'s ID and a value of $ v $'s ID in the out-neighbor list, and another with a key of $ v $'s ID and a value of $ u $'s ID in the in-neighbor list. For these two updates, {\graphkv} will adaptively select the optimal update method for each one, based on the threshold defined in Equation~\ref{eq: adaptive threshold}. Specifically, $d(u)$ in Equation~\ref{eq: C_P} and~\ref{eq: adaptive threshold} represents the total count of out-neighbors and in-neighbors of vertex $u$ in directed graphs. The degree sketch now also represents this total count, meaning that whenever vertex $u$ gains a new out-edge or in-edge, its degree sketch will increment by 1. The encoding method introduced in Section ~\ref{sec: encoding} can be conveniently extended to the directed graphs as well. Since the out-edge set and in-edge set are separately stored in the value of a certain entry while the vertex IDs are sorted inside each set. We can encode the out-edge set and in-edge set with the partitioned Elias-Fano workflow, respectively, and combine the results, maintaining the original layout of the value.

The primary focus of {\graphlsm} is to facilitate the efficient storing and processing of graph structure information (i.e., vertices and the edges adjacent to each vertex), which is the foundation of graph storage systems. 
At the same time, the adaptable nature of the key-value paradigm allows for the seamless incorporation of property graphs.
Specifically, in our implementation, we introduce an additional column family in RocksDB to store the mapping from vertices and edges to properties. The ID of the vertex or edge associated with the property key forms the key of the entry, which enables a fast property lookup by utilizing the prefix range lookup in the LSM-tree. 
User-defined secondary indexes can also be employed to speed up the query on vertices/edges with specific properties.

\newcommand{\vars}{\texttt}
\newcommand{\func}{\textrm}
\let\oldReturn\Return
\section{{\graphdb}: {\graphkv} Empowered Database}
\label{sec: graphdb}
This section introduces {\graphdb}, a comprehensive graph database that facilitates extensive graph storage applications, empowering efficient fundamental operations to intricate graph tasks. It enables users to perform queries through Gremlin with {\graphkv} as the storage backend. 
We will also introduce our distinctive graph traversal optimization for {\graphkv} and implementations of some necessary functions in graph database systems, such as transactions and multi-version concurrent control (MVCC).

\begin{figure*}[t]
\vspace{0mm}
    \centering    \includegraphics[width=0.95\textwidth]{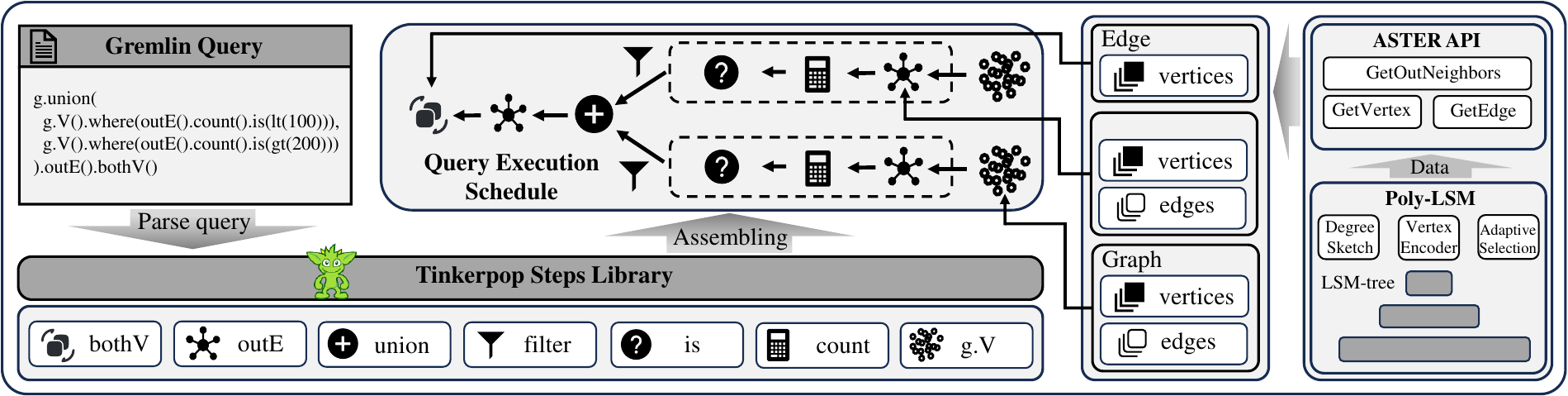}
    \vspace{-1mm}
    \caption{The architecture of {\graphdb}.}
    \vspace{-2mm}
    \label{fig: graphdb}
\end{figure*}

\vspace{0.5mm}
\noindent\textbf{Overview.}
The overall architecture of {\graphdb} is depicted in Figure ~\ref{fig: graphdb} which mainly comprises three fundamental components including graph query interface, query executor, and storage engine.
The graph query interface allows the users to precisely formulate various graph tasks on the graph data with straightforward statements. Additionally, with the integration of Gremlin language, this interface facilitates the convenient execution of these operations via the submission of Gremlin queries.
Upon reception of a Gremlin query, the query executor converts it into a viable execution schedule to execute through the APIs provided by the storage engine. To this end, {\graphdb} leverages the prominent graph computation framework, TinkerPop~\cite{tinkerpop}, to systematically break down the query statements into individual operations using a detailed step library. These operations are subsequently assembled to construct a query execution schedule as Figure ~\ref{fig: graphdb} presents. Then the schedule is executed sequentially with graph data accessed via the storage engine, the results of which are aggregated and returned to the user.

\vspace{0.5mm}
\noindent{\bf {\graphkv} Boosted Query Execution.}
As shown in Figure ~\ref{fig: graphdb}, a streaming computation graph is constructed when a Gremlin query is executed, which can be decomposed into a sequence of operations on graph data defined in the Tinkerpop step library. The execution of these operations depends on the interaction between the query executor and the underlying storage engine that manages the graph data. In this setup, {\graphdb} leverages {\graphkv} as its storage engine to ensure robust and efficient performance. {\graphkv} offers various interfaces, such as {\it AddVertex} and {\it GetOutNeighbors}, to facilitate the execution of these operations, thus enabling flexible manipulation of graph data and supporting various query execution plans for {\graphdb}. Furthermore, we tailor the Tinkerpop step library to our storage engine by implementing the {\it Vertex} and {\it Edge} interfaces. These interfaces access graph data through {\graphkv}, allowing the vertices and edges stored in {\graphkv} to be transformed into elements that can be processed by the computation graph, thereby completing the query execution process.

\vspace{0.5mm}
\noindent{\bf Query Optimization for {\graphkv}.} 
In addition to the specialized {\graphdb} interfaces, there are opportunities to further optimize query execution by leveraging the characteristics of the LSM-based storage engine. Specifically, {\graphdb} can utilize efficient range scans in the LSM-tree to avoid expensive random disk reads during a full graph scan (i.e., {\it g.V()}). Furthermore, the retrieval of properties and neighbors for a vertex or edge is deferred until necessary. Until then, {\graphkv} only verifies the existence of the vertex or edge and creates a placeholder with the appropriate ID.

\vspace{0.5mm}
\noindent{\bf Transaction and MVCC.} {\graphkv} is built on top of RocksDB~\cite{rocksdb}, which supports both pessimistic and optimistic transaction control. Given that write conflicts on the same key are rare in graph database systems, we employ the optimistic transaction policy by default. In this approach, conflict detection occurs when a transaction is committed, and the transaction will fail if there are conflicts with other writes. Read-write conflicts can also be escalated to write-write conflicts using the {\it GetForUpdate} interface. To ensure repeatable reads, {\graphkv} assigns a unique timestamp to each transaction and maintains it inline for each update of a graph element. Users can specify a snapshot with a timestamp using the {\it GetSnapshot} interface to obtain elements that are not newer than the specified timestamp. {\graphkv} ensures that elements visible to this snapshot are not reclaimed by compaction, allowing multi-version elements to coexist in the same LSM-tree.

\vspace{0.5mm}
\noindent{\bf Running Example.}
When a Gremlin query is issued to print vertices with out-edge counts less than 100 or greater than 200, {\graphdb} parses it into a bunch of elements, including graph components as vertexes and operations like edge lookups, using the Tinkerpop framework. These components are organized into a two-branch execution graph, where each branch retrieves vertices from {\graphkv} using the {\it GetOutNeighbors} interface and identifies vertices that meet the respective conditions. The results from both branches are then merged into a single set. Finally, the two vertices connected by the element’s outgoing edges are returned using the {\it GetEdge} and {\it GetVertex} interfaces.

\section{Related Works}
\noindent\textbf{Graph Database Systems.}
Over the past decade, extensive works have emerged on graph database research. Feng \etal \/ designed K\`{u}zu~\cite{feng2023kuzu} to optimize the join efficiency of graph database systems, which is further enhanced by Arroyuelo \etal~\cite{arroyuelo2024worst}.
TED~\cite{huang2023ted}, VisualNeo~\cite{huang2023visualneo}, and VINCENT~\cite{huang2022vincent} improve the subgraph enumeration and subgraph mining efficacy.
Luo \etal~\cite{luo2023multi} improved the performance of vertex-centric graph systems for large-scale distributed graph processing.
CGgraph~\cite{CuiLTY24} and MiniGraph~\cite{zhu2023minigraph} accelerate graph processing with CPU-GPU cooperative scheme and multi-core parallelism.
These approaches focus on graph processing tasks, specifically targeting the execution management and scheduling of complex graph algorithms~\cite{su2022banyan,fuchs2022sortledton}, instead of enhancing graph storage, thus being orthogonal to our work.
Additionally, since we focus on the generic graph database, works focusing on specific tasks are not closely related to ours, such as vector data storage~\cite{wang2024starling}, logic bugs identification~\cite{zhuang2023testing}, graph databased outsource ~\cite{chamani2023graphos}, scalable second-order random walk~\cite{li2022efficient} and temporal applications~\cite{hou2024aeong}.
 
The research on on-disk graph database for large-scale graphs ~\cite{kyrola2012graphchi, kyrola2014graphchi, freitas2020large, sakr2016large,li2022bytegraph} are more relevant to our work, thus the main-memory graph database~\cite{shi2024spruce,de2021teseo,chen2022g} are not examined.
GraphChi-DB~\cite{kyrola2014graphchi} is a disk-based system that leverages LSM-trees for efficient computation on large-scale graphs. GraphChi-DB adopts the partitioned adjacency list data structure from its well-known predecessor, GraphChi~\cite{kyrola2012graphchi}, which divides the complete adjacency list of a graph into several partitions on disk to reduce write amplification. However, we determined not to include GraphChi-DB in our experiential evaluation for 3 reasons: (1) GraphChi-DB is more akin to a storage engine rather than a complete graph database. It lacks support for running graph algorithms through a graph query language and does not incur the overhead of query parsing and execution. Therefore, directly comparing it with graph database systems like Neo4j would be unfair. (2) It was developed using Java 7, a very outdated version of Java, and is not compatible with later Java versions. (3) Its index structure has high read amplification, which limits its practical efficiency. The reason is that a vertex has its corresponding slice in each partition, it incurs at least one I/O at each partition to lookup any vertex's neighbors. 

Moreover, some graph processing systems tailor their designs to specific hardware characteristics and have presented impressive efficacy. For example, CAVE~\cite{papon2024cave} leverages the parallelism of SSD-based storage to achieve exceptional concurrency performance, while other studies~\cite{han2013turbograph,roy2013x,zhu2015gridgraph} address the challenge of slow random disk access, delivering sound results on traditional hard disk drives (HDDs). In contrast, our work focuses on a different task, graph storage optimization, with a more general solution that applies to any secondary storage device, thus placing it in a distinct scope.

\vspace{1mm}
\noindent\textbf{LSM-tree Optimization.}
LSM-tree is widely adopted as the storage backend for real-world key-value stores~\cite{rocksdb,polardb,tidb,cockroachdb,wiredtiger,lakshman2010cassandra}. Therefore, in recent years, there have been abundant studies that focused on optimizing LSM-trees in different aspects such as cost modeling ~\cite{dayan2017monkey, dostoevsky2018, liu2024structural, mo2023learning}, novel hardwares~\cite{huang2021nova, lu2017wisckey, yu2022treeline, chan2018hashkv}, and space efficiency~\cite{ren2017slimdb, mao2020comprehensive,sarkar2020lethe,dayanspooky,alkowaileet2019lsm}. 
To improve query efficiency, new filters ~\cite{dayan2021chucky,li2022seesaw}, data distribution conversion~\cite{zhang2022sa, zhang2022bi}, and spatial partitioning~\cite{vu2021incremental,eldawy2021beast} are introduced to reduce unnecessary data access.
For update performance enhancement, various index structures~\cite{sears2012blsm, raju2017pebblesdb, wu2015lsmtrie} and memory allocation strategies~\cite{luo2020breaking,kim2020robust, yu2024camal} are proposed.
Furthermore, due to the high adaptability and efficiency of the key-value paradigm, there is a trend toward applying LSM-trees in various other fields~\cite{shin2021lsm, chen2023chainkv, almodaresi2021incrementally, kim2017comparative, raju2018mlsm, qader2018comparative}. LSM-RUM-tree~\cite{shin2021lsm} proposed an effective structure that combines LSM-tree and R-tree to index spatial data. ChainKV~\cite{chen2023chainkv} leverage LSM-tree to store the blockchain of Ethereum systems. Almodaresi \etal~\cite{almodaresi2021incrementally} implemented an LSM-based system for answering sequence-level biological enquiries.
Our work focuses on optimizing LSM-tree under the scenario of graph storage. To the best of our knowledge, this is a novel problem that has yet to be scrutinized in existing works. 
Additionally, some LSM-based storage systems, such as RocksDB, support the \textit{Merge Operator} interface, which, as noted in~\cite{wu2022nebula,cao2020characterizing,xu2024ionia}, accumulates the values of two entries with the same key rather than obsoleting the older entry. This behavior is similar to how Poly-LSM handles delta entries and is utilized in our practical implementation. 

\section{Evaluation}
\label{sec: evaluation}
This section presents experimental evaluations of {\graphdb} against mainstream graph database systems under various workloads and tasks. The results demonstrate that {\graphdb} exhibits exceptional competitiveness relative to the baselines, excelling across a spectrum of fundamental graph queries and complex graph algorithms.

\subsection{Experimental Setup}

\noindent\textbf{Environment.}
Our experiments are conducted on a server with a 13th Gen Intel(R) Core(TM) i9-13900K CPU @ 4.0GHz processor, 128GB DDR4 main memory, and 2TB NVMe SSD, running 64-bit Ubuntu 20.04.4 LTS on an ext4 file system.

\vspace{0.5mm}
\noindent\textbf{Implementation.}
We implement {\graphkv} based on RocksDB~\cite{rocksdb}, a widely adopted LSM-based storage engine. We adhere to the default settings of RocksDB across most configurations, where $T=10$ and $B=4096$ bytes. The IDs of vertices are 64-bit integers. Furthermore, we set the bits-per-key of Bloom filters as 10 and the bits-per-vertex of degree sketch as 8. We implement {\graphdb} in JAVA, integrating the TinkerPop~\cite{tinkerpop} framework and connecting to {\graphkv} using Java Native Interface (JNI).

\vspace{0.5mm}
\noindent\textbf{Baselines.} 
We compare {\graphlsm} against existing LSM-tree structures for graph storage, specifically edge-based LSM-tree and vertex-based LSM-tree, designated as {\graphdbEB} and {\graphdbVB}. In addition, we also include {\graphlsm} without adaptive update mechanism, which trivially adopts either the delta update mechanism or the pivot update mechanism for all updates, designated as {\graphdbEBU} and {\graphdbVBU}. These baselines are also built on top of RocksDB, similar to {\graphkv}. In addition, most entries in {\graphdbVBU} exist in the form of edge-based adjacency lists. As a result, the performance of {\graphdbVB} and {\graphdbVBU} is nearly identical. Therefore, we only present the results for {\graphdbVB}.

We perform a comprehensive comparison of {\graphdb} against various mainstream graph database systems that include Neo4j (v3.2), ArangoDB (v2.8), OrientDB (v2.2), SQLG (PostgreSQL v9.6), JanusGraph (v1.0), and NebulaGraph (v3.8). Neo4j and OrientDB employ traditional linked list storage model while SQLG utilizes PostgreSQL~\cite{momjian2001postgresql} as its storage backend and represents a graph with relational models. JanusGraph utilizes a NoSQL storage engine BerkeleyDB~\cite{BerkeleyDB} as its backend by default, and maintains an adjacent list for each vertex. ArangoDB represents graph elements as documents and links them by indexes. NebulaGraph employs an edge-based LSM-tree that stores each edge as an entry.
These baselines all support the TinkerPop framework. To ensure fairness in evaluation, we use Gremlin as the graph query language for both {\graphdb} and all baseline databases, adhering to established practices in graph database benchmarking~\cite{microbenchmark}.

\begin{table}[t]
  \centering
  \small
\renewcommand\arraystretch{1.15}
\setlength{\tabcolsep}{3pt}
\scalebox{0.9}{
\hspace{-1.2mm}
  \begin{tabular}{@{}|l|l|r|r|r|r|@{}} \hline
    \multicolumn{1}{|c|}{\textbf{Scale}} & \multicolumn{1}{c|}{\textbf{Graph}}  & \multicolumn{1}{c|}{\bm{$n$}} & \multicolumn{1}{c|}{\bm{$m$}} & \multicolumn{1}{c|}{\bm{$\overline{d}$}} &  \multicolumn{1}{c|}{\bm{{Size}}} \\
    \hline
\multirow{4}{*}{Moderate} 
    & {DBLP}~\cite{yang2012defining}      & 317,080 & 1,049,866 & 3.31 & {131MB}\\     
    & {Twitch}~\cite{rozemberczki2021twitch} & 168,114 & 6,797,557 & 40.43 & {213MB}    \\
    & {LDBC Datagen*~\cite{microbenchmark}} & {184,329} & {767,894} & {4.17} &{22MB}\\
    & {Wiki-Talk} & {2,394,385} & {5,021,410} & {2.10} &{140MB}\\
    & {Cit-Patents} & {3,774,768} & {16,518,947} & {4.38} &{351MB}\\
    \hline
\multirow{4}{*}{Large} 
    & {Wikipedia}~\cite{microbenchmark}      & 3,333,397 & 123,709,902 & 37.11 & {1.8GB} \\ 
   & {Orkut}~\cite{yang2012defining}      & 3,072,441 & 234,370,166  & 76.28 & {1.3GB} \\ 
   & {Freebase Large*~\cite{microbenchmark}}      & {28,408,172} & {31,475,362}  & {1.11} &{1.4GB}\\

   \hline
   {Massive} & {Twitter}~\cite{kwak2010twitter}     & 41,652,230 & 1,202,513,046 & 57.74 & {24GB} \\ 
    \hline
  \end{tabular}
  }
  \vspace{1.6mm}
 \caption{Datasets. $n$ is the total number of vertices; $m$ is the number of edges; $\bar{d}$ represents the average degree. The datasets with * are property graphs.}
 \vspace{-6mm}
 \label{table: datasets}
 \setlength{\textfloatsep}{0pt}
\end{table}

\vspace{0.5mm}
\noindent\textbf{Graph Datasets.}
We employ five real graph datasets that are commonly used in graph processing and benchmarking to assess the effectiveness of our proposed methods in performing updates and lookups on graph structures, as detailed in Table~\ref{table: datasets}. 
These datasets span across various scales. 
The moderate-scale dataset includes DBLP and Twitch which have hundreds of thousands of vertices and Twitch contains significantly more edges.
Wikipedia and Orkut are large-scale datasets featuring millions of vertices and Orkut has higher average degree. Twitter is a massive-scale dataset with over a billion edges. 
In addition, LDBC Datagen and Freebase Large are property graphs from Lissandra's microbenchmark~\cite{microbenchmark}, used to evaluate the performance of property queries.
Furthermore, we also evaluate the performance of graph algorithms in LDBC Graphalytics~\cite{ldbcgraphalytics} with two LDBC-recommended datasets, Wiki-Talk and Cit-Patents.
We load these graphs into a graph database by sequentially inserting all edges in our experiments.

\vspace{0.5mm}
\noindent\textbf{Experiment Design.}
We evaluate the update and lookup performance of our proposed graph database, {\graphdb} by subjecting it to varied workloads encompassing different proportions of typical update and lookup operations, for comparison with the baseline graph database systems. Additionally, we assess the scalability of {\graphdb} across different-sized graph datasets like DBLP, Wikipedia, Orkut, and Twitter. 
For clarity, we focus on two representative operations: adding edges and retrieving neighbors. These operations represent common tasks in online graph database services.
Furthermore, we generalize the workloads by evaluating various graph queries, including graph structure updates, graph traversal, and property queries. Lastly, while analytical workloads like full graph scans are not our primary focus, we also demonstrate {\graphdb}'s potential in handling substantial graph scans.
Moreover, we also compare our core design, {\graphlsm}, against a variety of LSM-tree-based baselines across diverse workloads to scrutinize its efficacy.
We further validate the effectiveness of our cost model by comparing actual performance with our theoretical predictions.

\begin{figure*}[t]
    \centering  
    \hspace{4.5mm}
    \includegraphics[width=1.00\linewidth]{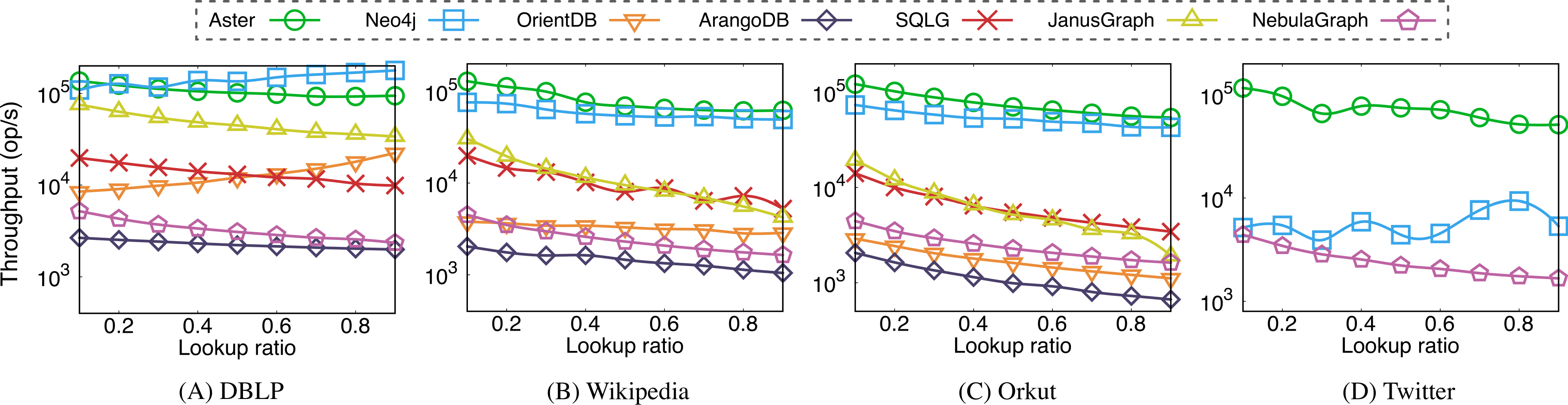}
    \vspace{-4mm}
\caption{{\graphdb} demonstrates strong efficiency and scalability across various workloads and dominates all baselines on large-scale and massive-scale graph datasets.}
\vspace{-4mm}
\label{fig: real world}
\end{figure*}

\subsection{System Performance}
\label{sec: sys_performance}
\noindent\textbf{{\graphdb} outperforms mainstream graph database systems in diverse workloads and demonstrates better scalability.} In Figure~\ref{fig: real world}, we compare the throughput of {\graphdb} against real-world graph database systems across a spectrum of workloads, ranging from write-heavy tasks (10\% lookups and 90\% updates) to read-heavy tasks (90\% lookups and 10\% updates), on various graph datasets from moderate-scale to massive-scale.
On DBLP, Neo4j and {\graphdb} demonstrates strong performance, surpassing other graph database systems, while {\graphdb} shows slightly slower performance with heavy lookups. 
The slowdown of {\graphdb} is due to the relatively small scale of the DBLP dataset, causing the LSM-tree in {\graphlsm} to have fewer levels. Consequently, the immutable table in Level 0 would markedly lower the overall read performance.
In contrast, Neo4j uses link lists as its backbone, whereas querying neighbors in a smaller data scale involves fewer random jumps, resulting in superior throughput capabilities. However, as the graph size increases, the limitations of the linked list become evident.
Specifically, both Neo4j and OrientDB use an adjacent linked list storage model, where each vertex maintains a head pointer to its linked list of edges. Since the linked list is not stored contiguously on disk, retrieving all edges for a vertex requires multiple I/O operations. This issue becomes more significant for high-degree vertices, leading to performance degradation on dense datasets such as Wikipedia and Orkut, as shown in Figure ~\ref{fig: real world}(B) to (C).

The relational-table-based baseline, SQLG, is slower than linked-list-based baselines due to the significant overhead involved in traversal using join operations, which scans through the vertex table and edge table incurring substantial I/O cost. This overhead results in performance that is roughly 10 to 100 times slower than Neo4j.
JanusGraph's neighbor retrieval is inefficient because BerkeleyDB stores pointers to key-value entries in its leaf nodes, leading to significant random disk access when retrieving multiple edges.
Consequently, JanusGraph shares the same drawbacks as linked-list-based storage and experiences performance degradation as the graph scale increases, as shown in Figure ~\ref{fig: real world}(B) and (C).
ArangoDB does not perform well in this experiment because it uses traditional document mapping storage, which leads to significant write amplification. Moreover, the necessity for frequent random accesses to vertices and edges documents also incurs substantial I/O costs.

Benefiting from the use of the LSM-tree structure, NebulaGraph exhibits substantial scalability, maintaining robust performance on the Twitter dataset without the significant degradation seen in many other baselines. Nevertheless, its performance on moderate-scale datasets is relatively mediocre. This is because Nebula utilizes an edge-based storage model, enabling lightweight updates but incurring significant overhead in neighbor retrieval, as discussed in Section ~\ref{sec: basic layouts}. 
In addition, it is important to note that Nebula primarily optimizes for horizontal distribution, therefore its performance is less competitive on a single machine.

\begin{figure*}[t]
\vspace{0mm}
    \centering  
    \hspace{4.5mm}
    \includegraphics[width=0.80\linewidth]{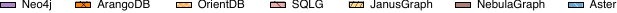}\\
    \vspace{-1.6mm}
    \hspace{-3mm}
    \includegraphics[width=1.00\linewidth]{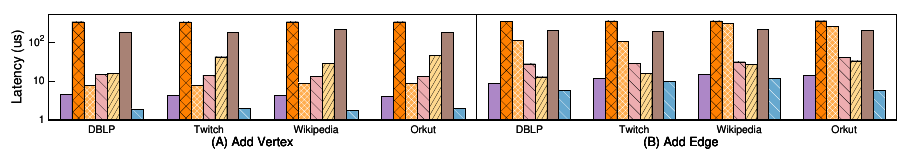}\\
    \vspace{-1.2mm}
    \hspace{-3mm}
    \includegraphics[width=1.00\linewidth]{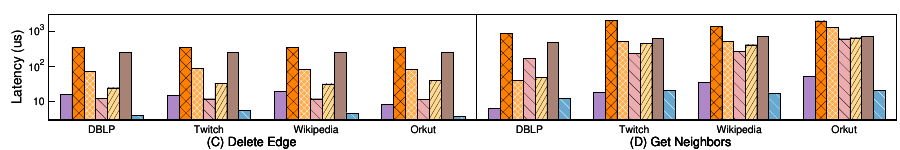}\\
    \vspace{-1.2mm}
    \hspace{-3mm}
    \includegraphics[width=1.00\linewidth]{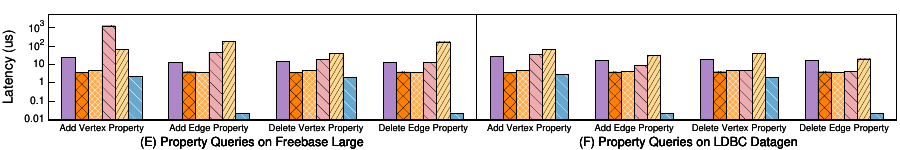}\\
    \vspace{-1.2mm}
    \hspace{-3mm}
    \includegraphics[width=1.00\linewidth]{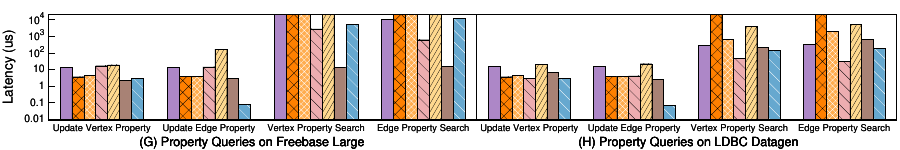}\\
\vspace{-2mm}
\caption{Time required for processing graph structure update, traversal, and property queries.}
\vspace{-2mm}
\label{fig: bar chart and ranking}
\end{figure*}

\begin{table}[t]
\vspace{1mm}
\small
\hspace{-1.5mm}
\makebox[0pt]{
\addtolength{\tabcolsep}{-2.7pt}
\renewcommand{\arraystretch}{1.15}
\scalebox{1}{
\begin{tabular}{|c|c|c|c|c|r|r|}
\hline
\multicolumn{1}{|c|}{System} & \multicolumn{1}{c|}{Add vertex} & \multicolumn{1}{c|}{Add edge} & \multicolumn{1}{c|}{Delete edge} & \multicolumn{1}{c|}{Get neighbor}  \\ \hline
Neo4j & 4.21us & 171.47us & 10.97us & 1302.51us\\ \hline
\graphdb & 1.79us & 6.03us & 3.87us & 20.67us \\ \hline
{NebulaGraph} & {238.14us} & {218.85us} & {220.04us} & {844.17us} \\ \hline
\end{tabular}
}
}
\vspace{0.6mm}
\caption{Latency of various graph queries on Twitter.}
\vspace{-7mm}
\label{table: twitter}
\end{table}

In contrast, {\graphdb} maintains satisfactory performance and outperforms all other databases on moderate and large-scale graphs. This is because the LSM-tree reduces write amplification by buffering incoming updates and flushing them sequentially. Additionally, the adaptive selection scheme ensures performance stability as the workload shifts. Moreover, the performance does not degrade with increasing graph scale, as shown in Figure ~\ref{fig: real world} (B) to (C). This stability is attributed to the fact that the overhead of edge retrieval and updates is not linearly proportional to the graph scale, resulting in a less significant increase in I/O as the graph scale grows, as analyzed in Section ~\ref{sec: adaptive}.

To further compare the scalability of these methods, we evaluate them on Twitter, a massive billion-scale dataset. With the data loading time limited to two days, all baselines except for Neo4j and {\graphdb} failed to complete the data loading process. In this experiment, as shown in Figure ~\ref{fig: real world} (D), the linked-list-based baseline, Neo4j, experienced continuous deterioration, with its throughput dropping nearly tenfold. In contrast, {\graphdb} maintained the same magnitude of throughput as observed in Figure ~\ref{fig: real world} (A) to (D).

\vspace{1mm}
\noindent\textbf{{\graphdb} consistently delivers efficient performance encompassing various graph operations.}
In Figure~\ref{fig: bar chart and ranking}, we provide a comparative study of {\graphdb} and all baseline models by reporting the average latency of executing a series of graph queries including fundamental graph traversal queries, graph structure manipulations and property-related graph queries across all datasets. Our experiments encompass the most commonly used operations in graph database systems, including adding vertex, adding edge, removing edge, getting neighbors, adding property, removing property, updating property, and getting edges/vertices according to given properties.
We report results from all moderate-scale and large-scale datasets in Figure~\ref{fig: bar chart and ranking}. All baselines apart from Neo4j and NebulaGraph time out while loading the massive-scale Twitter dataset. Therefore, we report the results for Twitter separately in Table ~\ref{table: twitter}.

As illustrated in Figure~\ref{fig: bar chart and ranking} (A)~$\sim$~~(C), {\graphdb} consistently achieves the lowest latency for all kinds of graph structure updates across various datasets, including the addition and removal of vertices and edges, benefiting from the efficient update of {\graphlsm}.
For the get neighbor operation in Figure~\ref{fig: bar chart and ranking} (D), {\graphdb}'s performance slightly trails Neo4j on moderate-scale datasets. In contrast, on large-scale datasets, {\graphdb} outperforms all baselines on these queries, demonstrating superior scalability for lookup-related queries.
The reason is that, as the data scale expands, the random access required by the linked-list-based structure increases significantly. In contrast, the number of read I/O operations for {\graphdb} grows more gradually and can be bounded by the number of levels in {\graphlsm}.

Figure~\ref{fig: bar chart and ranking} (E)~$\sim$~~(H) showcases the evaluation of graph property queries on our property graph datasets Freebase Large and LDBC Datagen.~\footnote{NebulaGraph requires a fixed schema for each vertex and edge, which prevents it from dynamically adding or deleting individual properties for a vertex or edge. Instead, it only supports updating all properties of an instance as a whole. As a result, NebulaGraph was not included in the results for Figure~\ref{fig: bar chart and ranking} (E) and (F).} The results indicate that our method excels in updates, such as adding, updating, and removing properties, outperforming all other baselines. Additionally, we demonstrate acceptable performance in searching vertices and edges by property, which is slightly slower than SQLG and Nebula. 
This is because both SQLG and Nebula stores edges or vertices in a continuous manner and filter them using a simple “where” clause, which avoids the need for join or random I/O operations. By employing appropriate indexes, their structures retrieve results relatively quickly.

\begin{figure*}[t]
    \vspace{0mm}
    \setlength{\belowcaptionskip}{-0.4cm}
    \begin{minipage}[b]{.7\linewidth}
    \hspace{-2.5mm}
    \includegraphics[width=\linewidth]{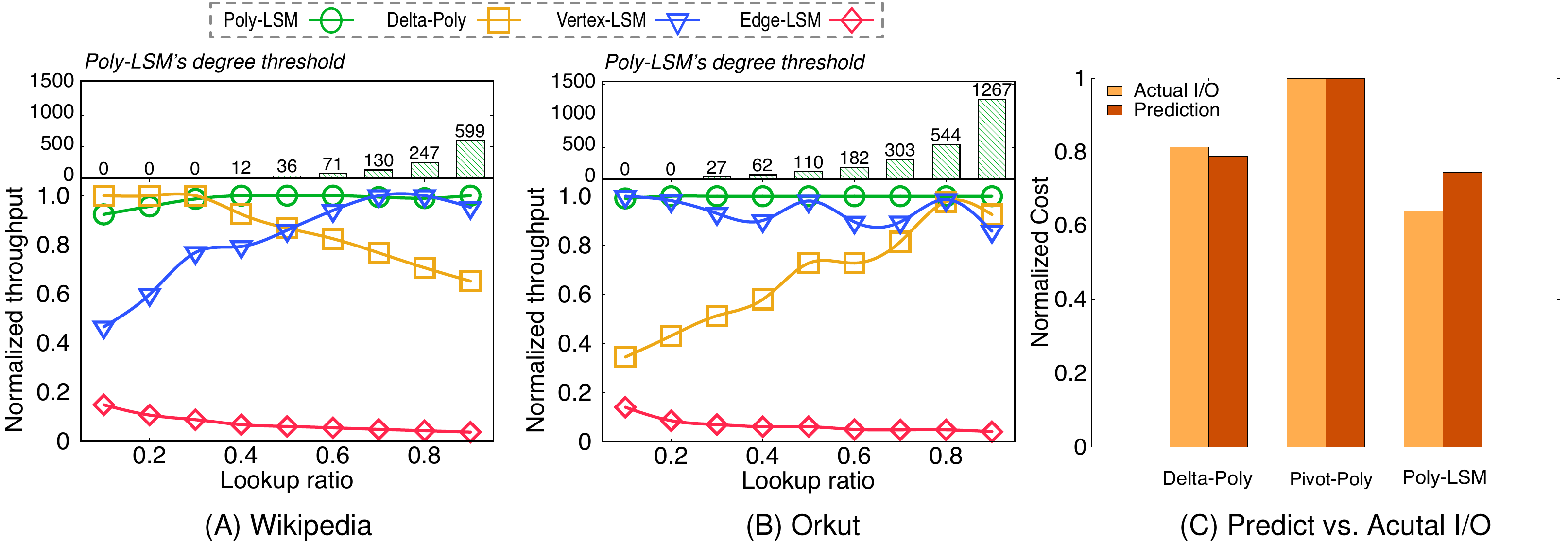}
    \end{minipage}%
\begin{minipage}[b]
{.3\linewidth}
\vspace{2.5mm}
\captionsetup[table]{singlelinecheck = false, labelfont={tiny, bf},
      textfont={tiny, bf}, justification=raggedright, margin={2mm, 1mm}, width=1.1\linewidth}
\captionof{table}{The statistics of {\graphdb} on Wikipedia under 50\% lookup ratio workload.
}
\vspace{-2.5mm}
\label{tab:IO comparison}
\vspace{-1mm}
\raisebox{17mm}{
\hspace{-3.5mm}
\renewcommand{\arraystretch}{1.1}
\scalebox{0.55}{
\begin{tabular}{|ll|}
\hline
\multicolumn{2}{|c|}{Statistics of Poly-LSM}                              \\ \hline
\multicolumn{1}{|l|}{CPU Time of Adaptive Mechanism} & 0.723 us / op     \\
\multicolumn{1}{|l|}{Morris Counter Read}                  & 0.107 us / op   \\
\multicolumn{1}{|l|}{Read Block Latency}                   & 10345.114 ms \\
\multicolumn{1}{|l|}{Write Block Latency}                  & 572.054 ms   \\
\multicolumn{1}{|l|}{Number of Vertex-based Updates}              & 879286       \\
\multicolumn{1}{|l|}{Number of Edge-based Updates}                & 1120714      \\
\multicolumn{1}{|l|}{Avg. Degree of Vertex-based Update}   & 3.58         \\
\multicolumn{1}{|l|}{Avg. Degree of Edge-based Update}     & 118.36       \\ \hline
\end{tabular}
}
}
\end{minipage}
\vspace{-7mm}
\caption{(A) and (B) shows {\graphlsm} exhibits strong robustness as the workload changes and outperforms other LSM-tree-based baselines. (C) indicates the cost model of {\graphlsm} accurately predicts the I/O overhead.}
\vspace{2mm}
\label{fig: poly lsm exp}
\end{figure*}

\vspace{1mm}
\noindent\textbf{{\graphkv} excels over other LSM-based graph storage designs.}
Figure~\ref{fig: poly lsm exp} (A) and (B) compared {\graphkv} against various LSM-tree-based baselines including edge-based LSM-tree (\graphdbEB), vertex-based LSM-tree (\graphdbVB), and {\graphlsm} using only delta updates (\graphdbEBU). We present the normalized throughput on our two large-scale graph datasets, Wikipedia and Orkut, under workloads consistent with the precedent experiment in Figure~\ref{fig: real world}. In addition, the top subfigures display the degree thresholds for {\graphkv}'s adaptive update mechanism under each corresponding workload, as specified in Equation~\ref{eq: d_t}.

{\graphdbVB} performs well for read-heavy scenarios but struggles under intensive graph updates since the read-and-modify scheme increases the cost of update while reducing the cost of lookup. In contrast, the performance of {\graphdbEBU} and {\graphdbEB} noticeably declines as the lookup ratio increases. Despite both employing edge-based update methods, the hybrid layout of {\graphlsm} enables {\graphdbEBU} to dominate {\graphdbEB} across all workloads, because the lookup performance of {\graphdbEB} is significantly hindered by its edge-based storage layout. Meanwhile, {\graphkv} outperforms all baselines that maintain optimal or near-optimal throughput across all workloads by adaptively selecting the best policy for each updated vertex, which has been proven to be theoretically global optimal in Section ~\ref{sec: adaptive}.

For instance, in the experiment on Wikipedia under a 50\% lookup ratio, {\graphlsm} outperforms both {\graphdbEBU} and {\graphdbVBU} by intelligently selecting either of them for each updated vertex based on its specific degree. The detailed statistics and the predictions of our model are shown in Figure ~\ref{fig: poly lsm exp} (C) and Table ~\ref{tab:IO comparison}. The predicted cost in Figure ~\ref{fig: poly lsm exp} (C) closely matches the actual cost, indicating that our model accurately captures the cost of different update methods. 
Additionally, we provide a detailed latency breakdown in Table ~\ref{tab:IO comparison} to support a more comprehensive evaluation. As shown, the edge count estimation and adaptive mechanism determination take just 0.1 µs and 0.7 µs, respectively, which contributes a small portion of the overall overhead. Hence we can achieve obviously enhanced system performance with minimal and manageable additional CPU overhead which further validates the effectiveness of our method. Moreover, the selection of pivot or delta updates follows our expectations, with pivot updates being applied to vertices with higher degrees and delta updates to those with lower degrees.

\begin{figure}
    \centering
    \includegraphics[width=.25\linewidth]{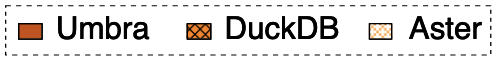}\\
    \vspace{-0.4mm}
    \includegraphics[width=.6\linewidth]{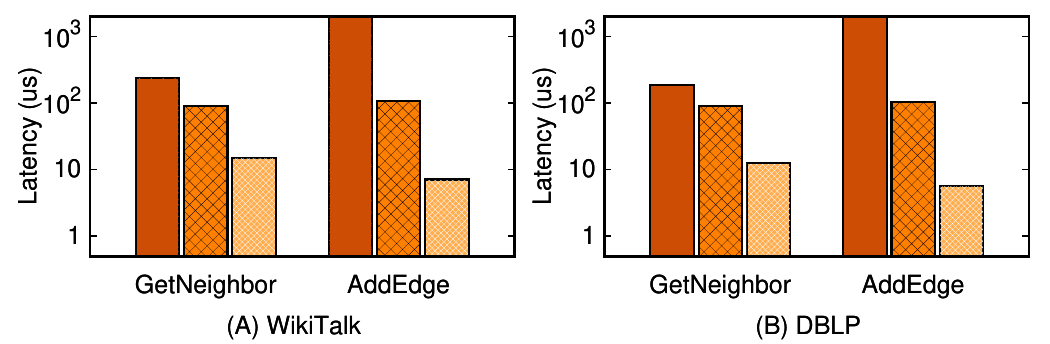}
    \vspace{-3mm}
    \caption{{In-memory comparison with memory-centric systems.}}
    \vspace{-4mm}
    \label{fig:inmemory}
\end{figure}
\vspace{1mm}
\noindent{\bf Competitive performance on small dataset.}
To demonstrate the robustness of {\graphdb} on small datasets that fit into memory, we conducted a case study evaluating
performance on two datasets, WikiTalk and DBLP, against two memory-centric storage systems, DuckDB~\cite{raasveldt2019duckdb} and Umbra~\cite{neumann2020umbra}. The results in Figure~\ref{fig:inmemory} indicate that {\graphdb} outperforms both baselines in the basic graph traversal operation, {\it GetNeighbors}, and the graph update operation, {\it AddEdge}. 
The main reason is that DuckDB and Umbra are not graph-structure-aware systems designed specifically for graph storage, like Aster.
For example, DuckDB processes analytical SQL queries efficiently by employing a relational data model, yet this model is less effective for graph traversal.
Additionally, Umbra’s use of sophisticated SQL compilation techniques, Just-In-Time (JIT) compilation, which convert queries into lightweight Intermediate Representation (IR) and then compile them to executable binary code, provides significant performance benefits for complex queries on large datasets but introduces overhead in simpler queries and smaller datasets, leading to weaker performance in this experiment.

\begin{table}[t]
\scalebox{0.75}{
\begin{tabular}{|c|rrrrr|rrrrr|}
\hline
Dataset &
  \multicolumn{5}{c|}{Cit-Patents ({\bf n}=3,774,768, {\bf m}=16,518,947)} &
  \multicolumn{5}{c|}{Wiki-Talk ({\bf n}=2,394,385, {\bf m}=5,021,410)} \\ \hline
Algorithm &
  \multicolumn{1}{c}{PageRank} &
  \multicolumn{1}{c}{CDLP} &
  \multicolumn{1}{c}{WCC} &
  \multicolumn{1}{c}{SSSP} &
  \multicolumn{1}{c|}{BFS} &
  \multicolumn{1}{c}{PageRank} &
  \multicolumn{1}{c}{CDLP} &
  \multicolumn{1}{c}{WCC} &
  \multicolumn{1}{c}{SSSP} &
  \multicolumn{1}{c|}{BFS} \\ \hline
Aster &
  491.886 &
  \underline{\textbf{334.991}} &
  1187.583 &
  \underline{\textbf{0.013}} &
  \underline{\textbf{0.019}} &
  287.229 &
  \underline{\textbf{115.912}} &
  267.540 &
  \underline{\textbf{0.118}} &
  11.638 \\
Neo4j &
  \underline{\textbf{294.540}} &
  479.899 &
  \underline{\textbf{684.763}} &
  0.038 &
  0.046 &
  \underline{\textbf{127.642}} &
  176.790 &
  \underline{\textbf{117.614}} &
  0.300 &
  \underline{\textbf{9.726}} \\
OrientDB &
  1604.027 &
  3211.459 &
  5246.498 &
  0.053 &
  0.051 &
  441.791 &
  910.428 &
  718.166 &
  37.536 &
  36.372 \\
SQLG &
  2626.018 &
  5964.497 &
  timeout &
  1.290 &
  1.424 &
  1198.875 &
  3651.953 &
  1862.763 &
  3.244 &
  5676.086 \\
JanusGraph &
  1593.294 &
  1912.659 &
  3029.494 &
  0.302 &
  0.419 &
  563.823 &
  657.349 &
  480.889 &
  0.687 &
  timeout \\
ArangoDB &
  timeout &
  timeout &
  timeout &
  2.257 &
  2.272 &
  timeout &
  timeout &
  timeout &
  1052.341 &
  1039.346 \\ \hline
  {GridGraph} &
  {0.570} &
    {7.430} &
    {0.410} &
    {0.380} &
    {0.360} &
    {0.300} &
    {3.210} &
    {0.130} &
    {0.040} &
    {0.250} \\
  {Mosaic} &
    {19.40} &
    {127.68} &
    {3.720} &
    {1.690} &
    {1.590} &
    {3.227}&
    {32.10} &
    {2.019} &
    {3.994} &
    {1.656} \\ \hline
\end{tabular}
}
\vspace{1mm}
\caption{This table presents the latency (in seconds) of {\graphdb} compared to various baselines in the LDBC Graphalytics Benchmark. The baselines are categorized into two groups: graph databases (top) and graph processing systems (bottom). While {\graphdb} is slower than the graph processing systems, it demonstrates solid performance relative to other graph databases. }
\vspace{-4mm}
\label{tab:graphalytics}
\end{table}

\vspace{1mm}
\noindent\textbf{{\graphkv} achieves desirable performance on LDBC Graphalytics benchmark.} 
In Table~\ref{tab:graphalytics}, we evaluate {\graphdb} and other graph database baselines using the LDBC Graphalytics benchmark~\cite{ldbcgraphalytics}.
Although substantial analytical algorithms are typically not the workload for online graph database systems and are usually conducted on OLAP graph processing systems such as GraphScope~\cite{fan2021graphscope}, we still aim to demonstrate the potential of {\graphlsm} to adapt to analytical workloads. 
We test five algorithms specified in the LDBC benchmark on two LDBC datasets, including PageRank, Community Detection using Label Propagation (CDLP), Weakly Connected Component (WCC), Single-Source Shortest Path (SSSP), and Breadth-First Search (BFS). In addition, we also compare these graph databases with two representative graph processing systems, GridGraph~\cite{zhu2015gridgraph} and Mosaic~\cite{mosaic}, to demonstrate the gap in processing analytical workloads for graph databases.
The time limit for conducting these algorithms is set to 2 hours. As shown in Table~\ref{tab:graphalytics}, {\graphdb} and Neo4j outperform other baselines remarkably. Generally, Neo4j performs better on algorithms requiring full graph scans, such as PageRank and WCC, while {\graphdb} shows strengths in algorithms that require local graph traversal, such as BFS and SSSP. Notably, both graph processing systems outperform all graph databases by a factor of 10 to 100, particularly in PageRank, CDLP\footnote{Note that CDLP requires tracking the number of labels for each vertex's neighbors, which is not well-suited to the edge-centric co-processing model used in Mosaic and GridGraph, resulting in relatively slower performance.}, and WCC, which require extensive full graph traversal. However, for BFS and SSSP, certain graph databases like {\graphdb} and Neo4j outperform Mosaic and GridGraph on the Cit-Patents dataset. This is likely due to the latter's reliance on substantial preprocessing and concurrent scheduling, which can be time-consuming and less effective for simpler queries. This result highlights the need for graph databases to improve their support for full graph traversal, suggesting a promising direction for future development.

\section{Conclusion}
We introduce {\graphkv}, a high-performance LSM-tree-based storage engine tailored for large-scale graphs. It achieves commendable performance in both update and lookup operations, leveraging a hybrid graph storage model, adaptive edge update mechanism, and refined entry encoding strategy. Based on {\graphkv}, we propose a robust and versatile graph database, {\graphdb}, that outperforms mainstream graph databases in terms of graph structure storage.

\bibliographystyle{ACM-Reference-Format}
\bibliography{main}

\end{document}